\documentclass[%
 reprint,
superscriptaddress,
 amsmath,amssymb,
 aps,
 prc,
]{revtex4-2}

\usepackage{graphicx}% Include figure files
\usepackage{dcolumn}% Align table columns on decimal point
\usepackage{bm}% bold math
\usepackage{xcolor}
\usepackage{comment}
\usepackage{multirow,bigdelim}

\usepackage{mathtools}
\usepackage{tabularx}

\usepackage{tikz}

\definecolor{darkgreen}{rgb}{0.0, 0.6, 0.0}
\tikzstyle{very densely dashed 1}=          [dash pattern=on 1.7pt off 1.136125pt]
\tikzstyle{very densely dashed 2}=          [dash pattern=on 1.7pt off 0.898pt]
\tikzstyle{very densely dashed 3}=          [dash pattern=on 1.5pt off 0.92pt] %

\DeclarePairedDelimiter\abs{\lvert}{\rvert}%
\DeclarePairedDelimiter\norm{\lVert}{\rVert}%

\usepackage{tikz}
\def\checkmark{\tikz\fill[scale=0.4](0,.35) -- (.25,0) -- (1,.7) -- (.25,.15) -- cycle;} 

\makeatletter
\let\oldabs\abs
\def\abs{\@ifstar{\oldabs}{\oldabs*}}
\let\oldnorm\norm
\def\norm{\@ifstar{\oldnorm}{\oldnorm*}}
\makeatother

\mathchardef\mhyphen="2D % Define a "math hyphen"

\newcommand{\beam}{\textrm{\fontsize{7}{8}\selectfont beam}}

\newcommand{\goodchi}{\protect\raisebox{2pt}{$\chi$}}

\makeatletter
\def\bbordermatrix#1{\begingroup \m@th
  \global\let\perhaps@scriptstyle\scriptstyle
  \@tempdima 4.75\p@
  \setbox\z@\vbox{%
    \def\cr{%
      \crcr
      \noalign{%
        \kern2\p@
        \global\let\cr\endline
        \global\let\perhaps@scriptstyle\relax
      }%
    }%
    \ialign{$\make@scriptstyle{##}$\hfil\kern2\p@\kern\@tempdima
      &\thinspace\hfil$\perhaps@scriptstyle##$\hfil
      &&\quad\hfil$\perhaps@scriptstyle##$\hfil\crcr
      \omit\strut\hfil\crcr
      \noalign{\kern-\baselineskip}%
      #1\crcr\omit\strut\cr}}%
  \setbox\tw@\vbox{\unvcopy\z@\global\setbox\@ne\lastbox}%
  \setbox\tw@\hbox{\unhbox\@ne\unskip\global\setbox\@ne\lastbox}%
  \setbox\tw@\hbox{$\kern\wd\@ne\kern-\@tempdima\left[\kern-\wd\@ne
    \global\setbox\@ne\vbox{\box\@ne\kern2\p@}%
    \vcenter{\kern-\ht\@ne\unvbox\z@\kern-\baselineskip}\,\right]$}%
  \null\;\vbox{\kern\ht\@ne\box\tw@}\endgroup}
\def\make@scriptstyle#1{\vcenter{\hbox{$\scriptstyle#1$}}}
\makeatother

\makeatletter
\def\pbordermatrix#1{\begingroup \m@th
  \global\let\perhaps@scriptstyle\scriptstyle
  \@tempdima 4.75\p@
  \setbox\z@\vbox{%
    \def\cr{%
      \crcr
      \noalign{%
        \kern2\p@
        \global\let\cr\endline
        \global\let\perhaps@scriptstyle\relax
      }%
    }%
    \ialign{$\make@scriptstyle{##}$\hfil\kern2\p@\kern\@tempdima
      &\thinspace\hfil$\perhaps@scriptstyle##$\hfil
      &&\quad\hfil$\perhaps@scriptstyle##$\hfil\crcr
      \omit\strut\hfil\crcr
      \noalign{\kern-\baselineskip}%
      #1\crcr\omit\strut\cr}}%
  \setbox\tw@\vbox{\unvcopy\z@\global\setbox\@ne\lastbox}%
  \setbox\tw@\hbox{\unhbox\@ne\unskip\global\setbox\@ne\lastbox}%
  \setbox\tw@\hbox{$\kern\wd\@ne\kern-\@tempdima\left(\kern-\wd\@ne
    \global\setbox\@ne\vbox{\box\@ne\kern2\p@}%
    \vcenter{\kern-\ht\@ne\unvbox\z@\kern-\baselineskip}\,\right)$}%
  \null\;\vbox{\kern\ht\@ne\box\tw@}\endgroup}
\def\make@scriptstyle#1{\vcenter{\hbox{$\scriptstyle#1$}}}
\makeatother

\makeatletter
\DeclareRobustCommand{\mypm}{\mathbin{\mathpalette\@mypm\relax}}
\DeclareRobustCommand{\@mypm}[2]{\ooalign{%
  \raisebox{.1\height}{$#1+$}\cr
  \smash{\raisebox{-.6\height}{$#1-$}}\cr}}
\makeatother

\definecolor{darkgrey}{rgb}{0.130416, 0.130416, 0.130416}

\usepackage{etoolbox}

\makeatletter
\appto\abstract{%
  \let\latexlist\list
  \def\list{\edef\keeprightskip{\the\rightskip}\latexlist}%
  \patchcmd\latexlist{\ignorespaces}{\rightskip\keeprightskip\ignorespaces}{}{}%
}
\makeatother

\begin{document}

\preprint{APS/123-QED}

%----------------------------
%   Possible titles
% \title{Chronicling the total width for the $3_{1}^{-}$ resonance in $\mathrm{^{12}C}$: \\the formal, the observed and the FWHM}
% \title{The chronicles of the $3_{1}^{-}$ total width in $\mathrm{^{12}C}$:\\ the formal, the observed and the FWHM}
\title{Understanding the total width of the $3_{1}^{-}$ state in $\mathrm{^{12}C}$}
\thanks{Formerly: The chronicles of the $3_{1}^{-}$ total width in $\mathrm{^{12}C}$:\\ the formal, the observed and the FWHM}

% \homepage[\phantomsection\label{lb:homepage}]{Formerly: The chronicles of the $3_{1}^{-}$ total width in $\mathrm{^{12}C}$:\\ the formal, the observed and the FWHM}

% \title{Manuscript Title:\\with Forced Linebreak}% Force line breaks with \\

\author{K.~C.~W.~Li}
% \altaffiliation{Present address: Department of Physics, University of Oslo, N-0316 Oslo, Norway}%
\email{k.c.w.li@fys.uio.no}
\affiliation{Department of Physics, University of Oslo, N-0316 Oslo, Norway}

\author{R.~Neveling}%
\affiliation{iThemba LABS, National Research Foundation, PO Box 722, Somerset West 7129, South Africa}

\author{P.~Adsley}%
\affiliation{Department of Physics and Astronomy, Texas A\&M University, College Station, Texas 77843, USA}
\affiliation{Cyclotron Institute, Texas A\&M University, College Station, Texas 77843, USA}

\author{H.~Fujita}%
\affiliation{Research Center for Nuclear Physics, Osaka University, Ibaraki, Osaka 567-0047, Japan}

\author{P.~Papka}%
\affiliation{iThemba LABS, National Research Foundation, PO Box 722, Somerset West 7129, South Africa}
\affiliation{Department of Physics, University of Stellenbosch, Private Bag X1, 7602 Matieland, South Africa}

\author{F.~D.~Smit}%
\affiliation{iThemba LABS, National Research Foundation, PO Box 722, Somerset West 7129, South Africa}

\author{J.~W.~Br\"ummer}%
% \author{\\J.~W.~Br\"ummer}%
\affiliation{iThemba LABS, National Research Foundation, PO Box 722, Somerset West 7129, South Africa}
\affiliation{Department of Physics, University of Stellenbosch, Private Bag X1, 7602 Matieland, South Africa}

\author{L.~M.~Donaldson}%
\affiliation{iThemba LABS, National Research Foundation, PO Box 722, Somerset West 7129, South Africa}
\affiliation{School of Physics, University of the Witwatersrand, Johannesburg 2050, South Africa}

% \author{M.~Freer}%
% \affiliation{School of Physics and Astronomy, University of Birmingham, Edgbaston, Birmingham, B15 2TT, United Kingdom}

\author{M.~N.~Harakeh}%
\affiliation{Nuclear Energy Group, ESRIG, University of Groningen, 9747 AA Groningen, The Netherlands}

\author{Tz.~Kokalova}%
\affiliation{School of Physics and Astronomy, University of Birmingham, Edgbaston, Birmingham, B15 2TT, United Kingdom}

\author{E.~Nikolskii}%
\affiliation{NRC Kurchatov Institute, Ru-123182 Moscow, Russia}

\author{W.~Paulsen}%
\affiliation{Department of Physics, University of Oslo, N-0316 Oslo, Norway}

\author{L.~Pellegri}%
\affiliation{iThemba LABS, National Research Foundation, PO Box 722, Somerset West 7129, South Africa}
\affiliation{School of Physics, University of the Witwatersrand, Johannesburg 2050, South Africa}

\author{S.~Siem}%
\affiliation{Department of Physics, University of Oslo, N-0316 Oslo, Norway}

\author{M.~Wiedeking}%
\affiliation{iThemba LABS, National Research Foundation, PO Box 722, Somerset West 7129, South Africa}
\affiliation{School of Physics, University of the Witwatersrand, Johannesburg 2050, South Africa}

\date{\today}% It is always \today, today,

\begin{abstract}
\begin{description}
%------------------------------------------------
\item[Background]
Recent measurements indicate that the previously established upper limit for the $\gamma$-decay branch of the $3_{1}^{-}$ resonance in $^{12}\textrm{C}$ at $E_{x} = 9.641(5)$ MeV may be incorrect.
As a result, the $3_{1}^{-}$ resonance has been suggested as a significant resonance for mediating the triple-$\alpha$ reaction at high temperatures above 2 GK.
Accurate estimations of the $3_{1}^{-}$ contribution to the triple-$\alpha$ reaction rate require accurate knowledge of not only the radiative width, but also the total width.
%------------------------------------------------
\item[Purpose]
In anticipation of future measurements to more accurately determine the $\gamma$-decay branch of the $3_{1}^{-}$ resonance, the objective of this work is to accurately determine the total width of the $3_{1}^{-}$ resonance.
%------------------------------------------------
\item[Method]
% An evaluation was performed on all previous results considered in the current ENSDF average for the $3_{1}^{-}$ total width of 46(3) keV.
An evaluation was performed on all previous results considered in the current ENSDF average of 46(3) keV for the physical total width (FWHM) of the $3_{1}^{-}$ resonance in $^{12}\textrm{C}$.
A new \textbf{R}-matrix analysis for the $3_{1}^{-}$ resonance was performed with a self-consistent, simultaneous fit of several high-resolution $\mathrm{^{12}C}$ excitation-energy spectra populated with direct reactions.
%------------------------------------------------
\item[Results]
The global analysis performed in this work yields a formal total width of $\Gamma(E_{r}) = 46(2)$ keV and an observed total width of $\Gamma_{\textrm{obs}}(E_{r}) = 38(2)$ keV for the $3_{1}^{-}$ resonance.
%------------------------------------------------
\item[Conclusions]
Significant unaccounted-for uncertainties and a misstated result were discovered in the previous results employed in the ENSDF for the physical (or observed) total width of the $3_{1}^{-}$ resonance.
These previously reported widths are fundamentally different quantities, leading to an invalid ENSDF average.
An observed total width of $\Gamma_{\textrm{obs}}(E_{r}) = 38(2)$ keV is recommended for the $3_{1}^{-}$ resonance in $\mathrm{^{12}C}$.
This observed total width should be employed for future evaluations of the observed total radiative width for the $3_{1}^{-}$ resonance and its contribution to the high-temperature triple-$\alpha$ reaction rate.
\end{description}

\end{abstract}

\maketitle

%\tableofcontents

%========================================================================================
\section{\label{sec:Introduction}INTRODUCTION}

% The structure of $^{12}\textrm{C}$ at excitation energies just above the Hoyle state has remained the focus of $\alpha$-clustering studies.

At high temperatures of above 2 GK, the triple-$\alpha$ reaction proceeds through resonances above the Hoyle state.
Such high-temperature conditions are significant for astrophysical environments such as the shock front of type II supernovae.
In this temperature region, there is significant uncertainty in the triple-$\alpha$ rate, owing to the complexities of disentangling the broad resonances which intrinsically overlap and interfere \cite{PhysRevC.80.044304, PhysRevC.81.024303, PhysRevC.84.054308, FREER20141, RevModPhys.90.035004, LI2022136928, PhysRevC.105.024308}.
In contrast, the triple-$\alpha$ reaction at medium temperatures (between 0.1 and 2 GK) proceeds almost exclusively through the narrow primary peak of the Hoyle state.
% The significance of correctly including the broad resonances above the Hoyle state can be observed in the revised triple-alpha rate by Fynbo \textit{et al.}\cite{HOUFynbo_2005}. 
In order to understand the significance of correctly including the broader resonances above the Hoyle state for the triple-$\alpha$ rate at high temperatures, consider the differences in the triple-$\alpha$ rates calculated by Angulo \textit{et al.} and Fynbo \textit{et al.} \cite{ANGULO19993, HOUFynbo_2005}.
At the time of publication of both Refs. \cite{ANGULO19993} and \cite{HOUFynbo_2005}, the $2_{2}^{+}$ resonance was not yet conclusively observed.
Consequently, the $2_{2}^{+}$ resonance was omitted in the revised rate \cite{HOUFynbo_2005}.
In contrast, the NACRE rate \cite{ANGULO19993} assumed the existence of the $2_{2}^{+}$ resonance at $E_{x}$ = 9.1 MeV with $\Gamma = 0.56$ MeV, yielding an increase in the triple-$\alpha$ rate (above $\approx6$ GK) by several orders of magnitude relative to Ref. \cite{HOUFynbo_2005}.
In the past few decades, the nuclear-physics community has invested significant effort in the search for the $2_{2}^{+}$ rotational state, culminating in its eventual identification \cite{MItoh_2004, PhysRevC.80.041303, PhysRevC.84.054308, PhysRevC.86.034320, PhysRevLett.110.152502}.
Whilst this state has now been firmly established to exist at $E_{x} \approx$ 9.9 MeV, its exact properties are still somewhat uncertain due to it being submerged under the broad, surrounding $0^{+}$ strength.

An even greater source of uncertainty exists in the triple-$\alpha$ rate at high temperatures above 2 GK, where the Gamow window enables the reaction to proceed through the $3_{1}^{-}$ resonance in $\mathrm{^{12}C}$. 
In the past, the contribution from the $3_{1}^{-}$ resonance has been largely neglected given the reported upper limit on the radiative branching ratio of $\Gamma_{\textrm{rad}}$/$\Gamma < 8.3 \times 10^{-7}$ (95\% C.L.) \cite{PhysRevC.10.909}. 
Recent measurements have indicated that this upper limit may be incorrect. 
The first indication of this possible error was reported in a study by Tsumura \textit{et al.} \cite{TSUMURA2021136283} which yielded a branching ratio of $\Gamma_{\textrm{rad}}$/$\Gamma = 1.3_{-1.1}^{+1.2} \times 10^{-6}$, though the resolution in Ref. \cite{TSUMURA2021136283} for the $3_{1}^{-}$ resonance was low ($\approx$800 keV FWHM) and the identification/quantification of the $3_{1}^{-}$ peak is difficult given the complex and significant background.
Specifically, it is the branching ratio for the $E1$ $\gamma$-ray transition between the $3_{1}^{-}$ and $2_{1}^{+}$ states which is expected to dominate the total $\gamma$ decay of the $3_{1}^{-}$ resonance; the probability of the $E3$ $\gamma$ decay ($3_{1}^{-} \rightarrow 0_{\textrm{g.s.}}^{+}$) being determined in Ref. \cite{TSUMURA2021136283} as $6.7(10) \times 10^{-9}$ using the associated 0.31(4) meV width from an $(e,e^{\prime})$ measurement \cite{CRANNELL1967152} and the ENSDF average for the total width of the $3_{1}^{-}$ resonance of $\Gamma = 46(3)$ keV \cite{KELLEY201771}.
Since 2017, the ENSDF average for the $3_{1}^{-}$ total width has employed Refs. \cite{PhysRev.104.1059, PhysRev.125.992, PhysRevC.86.064306, PhysRevC.87.057307}, yielding an uncertainty-weighted average of 46(3) keV \cite{KELLEY201771} (see Table \ref{tab:DataUsedForNNDCEvaluationOfWidth}).
%================================================================================================
\begin{table}[b]%The best place to locate the table environment is directly after its first reference in text
\caption{\label{tab:DataUsedForNNDCEvaluationOfWidth}%
The reported results for the $3_{1}^{-}$ resonance in $\mathrm{^{12}C}$ which are employed in the current ENSDF average \cite{KELLEY201771}.
The ENSDF definition for the listed total width is the full width at half maximum (FWHM) intensity for a resonance.
$\Gamma(E_{r})$ is the formal total width (see Eq. \ref{eq:FormalTotalWidth}).
$\Gamma_{\textrm{obs}}$ is the observed total width (see Eq. \ref{eq:ObservedTotalWidth}). 
$\Gamma_{\textrm{FWHM}}$ is the FWHM of the intrinsic lineshape.
Averaging the total widths reported in Refs. \cite{PhysRev.104.1059, PhysRev.125.992, PhysRevC.86.064306, PhysRevC.87.057307} yields an average of 45(3) keV: this minor discrepancy with the currently listed 46(3) keV is ``due to a rounding judgment'' \cite{JHKelley_private}.
Details of the uncertainty policy for the ENSDF can be found in Refs. \cite{ENSDF_guidelines, V_AveLib}).
}
\begin{ruledtabular}
%================================================
\begin{tabular}{c D{,}{}{3.3} c D{,}{}{3.3} D{,}{}{3.3}}
Ref.    & \multicolumn{1}{c}{Resolution\footnotemark[1]} & $\Gamma(E_{r})$ & \multicolumn{1}{c}{$\Gamma_{\textrm{obs}}(E_{r})$} & \multicolumn{1}{c}{$\Gamma_{\textrm{FWHM}}$} \\
         & \multicolumn{1}{c}{{[keV]}} & {[keV]} & \multicolumn{1}{c}{[keV]} & \multicolumn{1}{c}{[keV]} \\ [0.3ex] \hline \\ [-2.3ex]
\citeauthor{PhysRev.104.1059} \cite{PhysRev.104.1059}     &\multicolumn{1}{c}{---}&---&\multicolumn{1}{c}{---}&30,(8)  \\ [0.3ex] \hline \\ [-2.3ex]
\citeauthor{PhysRev.125.992} \cite{PhysRev.125.992}      & \multicolumn{1}{c}{$\approx40$} &---& \multicolumn{1}{c}{---} & 36,(6)\footnotemark[2] \\ [0.3ex] \hline \\ [-2.3ex]
% \citeauthor{PhysRevC.86.064306} \cite{PhysRevC.86.064306}   & 60\textrm{--}120,/55\textrm{--}85 &---& \multicolumn{1}{c}{---} & 43,(4) \\ [0.3ex] \hline \\ [-2.3ex]
\citeauthor{PhysRevC.86.064306} \cite{PhysRevC.86.064306}   & \multicolumn{1}{c}{60\textrm{--}120} &---& \multicolumn{1}{c}{---} & 43,(4) \\ 
       & \multicolumn{1}{c}{55\textrm{--}85} & & & \\ [0.3ex] \hline \\ [-2.3ex]
% \citeauthor{PhysRevC.87.057307} \cite{PhysRevC.87.057307}   & 54,(2) & 48(2) & 39,(4)\footnotemark[3] &39,(4)\footnotemark[3] \\
\citeauthor{PhysRevC.87.057307} \cite{PhysRevC.87.057307}   & 54,(2)\footnotemark[3] & 48(2) & \multicolumn{1}{c}{---} & \multicolumn{1}{c}{---} \\
\end{tabular}
%================================================
\end{ruledtabular}
\footnotetext[1]{Reported as FWHM.}
\footnotetext[2]{The abstract and body of Ref. \cite{PhysRev.125.992} inconsistently report 35(6) and 36(6) keV, respectively; the ENSDF employs the latter \cite{KELLEY201771}.}
\footnotetext[3]{Determined from Ref. \cite{PhysRevC.87.057307} which incorrectly reports the Gaussian standard deviation of $\sigma = 23(1)$ keV as the FWHM.}
% \footnotetext[2]{Determined from Ref. \cite{PhysRevC.87.057307} which incorrectly reports the Gaussian standard deviation of $\sigma = 23(1)$ keV as the FWHM.}
% \footnotetext[3]{Not reported in Ref. \cite{PhysRevC.87.057307}; converted from $\Gamma(E_{r})$ in this work.
% The reported uncertainties include systematic errors which account for the choice in background (see text for details).}
\end{table}
%================================================================================================
A subsequent result of $\Gamma_{\textrm{rad}}$/$\Gamma = 6.4(51) \times 10^{-5}$ was reported by Cardella \textit{et al.} \cite{PhysRevC.104.064315}, however, the resolution is also relatively low and only $\approx3$ counts corresponding to the $3_{1}^{-}$ resonance were observed.
Unfortunately, the uncertainties from both measurements are substantial and lead to large uncertainties in the associated contribution to the triple-$\alpha$ rate at high temperatures. 
However, what is clear is that if the order of magnitude of these results is correct, the previously established upper limit of $\Gamma_{\textrm{rad}}$/$\Gamma < 8.3 \times 10^{-7}$ (95\% C.L.) may be incorrect. 
The upward trend for the associated radiative branching ratio for the $3_{1}^{-}$ resonance \cite{TSUMURA2021136283, PhysRevC.104.064315} suggests that the $3_{1}^{-}$ contribution may not only be significant, but dominant at temperatures above 2 GK.
The significant uncertainties in Refs. \cite{TSUMURA2021136283, PhysRevC.104.064315} do not enable the $3_{1}^{-}$ contribution to be meaningfully constrained and new, more sensitive measurements are required.
Accurate estimations for the observed total radiative width of the $3_{1}^{-}$ resonance, as well as its contribution to the triple-$\alpha$ reaction rate, require accurate knowledge of not only the $\gamma$-decay branching ratio, but also the total width.
% In anticipation of future measurements to more accurately determine the $\gamma$-decay branch, the objective of this work is to provide a new analysis for the total width of the $3_{1}^{-}$ resonance for comparison with the current ENSDF average.
In anticipation of future measurements to more accurately determine the $\gamma$-decay branch, the primary objective of this work is to provide a new analysis for the total width of the $3_{1}^{-}$ resonance.
A secondary objective is to perform a meta-analysis on the previous results considered in the current ENSDF average for the $3_{1}^{-}$ total width.
% All previous results considered in the ENSDF average for the $3_{1}^{-}$ total width are summarized in Table \ref{tab:DataUsedForNNDCEvaluationOfWidth}.
% Here, the ENSDF definition for the listed total width is the full width at half maximum (FWHM) intensity for a resonance \cite{KELLEY201771}.

%================================================================================================
\section{\label{sec:DataAnalysisAndResults}DATA ANALYSIS AND RESULTS}

In this work, the primary analysis considers inclusive excitation-energy spectra from six different measurements (see Table \ref{tab:SummaryOfExperimentalParameters}).
% A portion of this data corresponds to that studied in a previous investigation for the predicted breathing-mode excitation of the Hoyle state \cite{LI2022136928, PhysRevC.105.024308}.
% In addition, a subset of the inelastic proton-scattering data employed in a previous study of the $3_{1}^{-}$ total width (the latest considered in the current ENSDF average) is also incorporated into this analysis \cite{PhysRevC.87.057307}.
The $\mathrm{^{12}C}(\alpha, \alpha^{\prime})\mathrm{^{12}C}$ and $\mathrm{^{14}C}(p, t)\mathrm{^{12}C}$ data have been previously employed in a previous investigation for the predicted breathing-mode excitation of the Hoyle state \cite{LI2022136928, PhysRevC.105.024308}.
The $\mathrm{^{12}C}(p, p^{\prime})\mathrm{^{12}C}$ spectrum studied in this work is an independently analysed subset of the data employed in a previous study of the $3_{1}^{-}$ total width \cite{PhysRevC.87.057307}.
These spectra were simultaneously fitted with phenomenological lineshape parameterizations from multi-level, multi-channel \textbf{R}-matrix theory.
The underlying formalism of the fit analysis is detailed in Section \ref{subsec:RMatrixFormalism} and
the details of this primary analysis are presented in Section \ref{subsec:The primary analysis of inclusive spectra}.
In addition, a meta-analysis was also performed on the previous studies considered on this current ENSDF average \cite{KELLEY201771}, with the exception of Ref. \cite{PhysRev.104.1059}.
These assessments are detailed in sections \ref{subsec:Assessing the $3_{1}^{-}$ total width from Kokalova et al.}, \ref{subsec:Assessing the $3_{1}^{-}$ total width from Alcorta et al.} and \ref{subsec:Assessing the $3_{1}^{-}$ total width from Browne et al.}.
A quantitative assessment of the $3_{1}^{-}$ total width reported in Ref. \cite{PhysRev.104.1059} was deemed unfeasible within the scope of this work as the methodology in Ref. \cite{PhysRev.104.1059} is significantly different from this work.
%========================================================================================
%   Table with slightly less information
\begin{table}[h]
\caption{\label{tab:SummaryOfExperimentalParameters}%
Summary of the experimental parameters for the inclusive excitation-energy spectra analysed in this work.
% The $\mathrm{^{12}C}(\alpha, \alpha^{\prime})\mathrm{^{12}C}$ and $\mathrm{^{14}C}(p, t)\mathrm{^{12}C}$ data have been previously used Refs. \cite{LI2022136928, PhysRevC.105.024308} and the $\mathrm{^{12}C}(p, p^{\prime})\mathrm{^{12}C}$ data corresponds to an independently analyzed subset of that from Ref. \cite{PhysRevC.87.057307}.
} 
\begin{ruledtabular}
\begin{tabular}{lD{.}{.}{2.0}ccc}
%---------
\multicolumn{1}{c}{Reaction} &
\multicolumn{1}{c}{Angle} &
\multicolumn{1}{c}{$E_{\beam}$} &
% \multicolumn{1}{c}{Full-acceptance\footnotemark[2]} &
\multicolumn{1}{c}{Target} &
\multicolumn{1}{c}{Fitted $E_{x}$} \\
% \\ [0.2ex] \cline{9-10} \\ [-2.3ex]
%---------
&
\multicolumn{1}{c}{[deg]} &
\multicolumn{1}{c}{[MeV]} &
($\mu$g/cm$^{2}$) &
range [MeV]
\vspace{2pt} \\
%---------
\hline \\ [-2.3ex]
$\mathrm{^{12}C}(\alpha, \alpha^{\prime})\mathrm{^{12}C}$   & 0  & 118  & \,\,\,$\mathrm{^{nat}C}$ (1053) & \hphantom{0}5.0--14.8 \\
                                                            & 0  & 160  & $\mathrm{^{nat}C}$ (300)  & \hphantom{0}7.3--20.0 \\
                                                            & 10 & 196  & $\mathrm{^{nat}C}$ (290)  & 7.15--21.5 \\
\hline \\ [-2.3ex]
$\mathrm{^{14}C}(p, t)\mathrm{^{12}C}$  & 0     & 100       & $\mathrm{^{14}C}$ (280)   & \hphantom{0}6.0--15.3 \\
                                        & 21    & 67.5      & $\mathrm{^{14}C}$ (300)   & \hphantom{0}6.8--14.5 \\ \hline \\ [-2.3ex]
$\mathrm{^{12}C}(p, p^{\prime})\mathrm{^{12}C}$  & 16     & 66       & $\mathrm{^{nat}C}$ (1000)   & \hphantom{0}6.0--15.3 \\
\end{tabular}
\end{ruledtabular}
% \footnotetext[1]{Only the charged-particle decays were analyzed as the Hoyle state was not fully accepted on the focal plane.}
\end{table}
%------------------------------------------------

%================================================================================================
\subsection{\label{subsec:RMatrixFormalism}\textbf{R}-matrix formalism}

A comprehensive description of the phenomenological \textbf{R}-matrix formalism for this analysis is given in Ref. \cite{PhysRevC.105.024308}, with the pertinent components described here. 
For this analysis of direct-reaction data, consider a direct reaction populating a recoil nucleus which subsequently decays, represented as
% A brief overview is here provided (see Ref. \cite{PhysRevC.105.024308} for a detailed overview).
% A brief overview of this formalism is here provided, with a more detailed overview in Ref. \cite{PhysRevC.105.024308}.
%------------------------------------------------
\begin{equation}
   \centering
	A + a \rightarrow B + b \;\;\;\;\;\; (B \rightarrow C + c),
   \label{eq:ResonancePopulation_NuclearReaction}
\end{equation}
%------------------------------------------------
for the target ($A$), projectile ($a$), recoil ($B$), ejectile ($b$) and decay products from the recoil ($C$ and $c$).
The intrinsic lineshape observed for excitations of a particular spin and parity is given by
% This is equivalent to a special case of full multilevel formalism for the intrinsic lineshape observed for excitations of a particular spin and parity, populated with direct reactions of the form:
%------------------------------------------------
%%%%	GeneralNLevelCrossSection
\begin{equation}
  \centering
	N_{ab,c}(E)  = P_{c} \abs{\, \sum\limits_{\lambda, \mu}^{N} G_{\lambda ab}^{\frac{1}{2}} \gamma_{\mu c} A_{\lambda \mu}}^{2},
  \label{eq:GeneralNLevelCrossSection}
\end{equation}
%------------------------------------------------
\noindent where $\gamma$ is the reduced-width amplitude and $A_{\lambda \mu}$ is an element of the level matrix.
Subscript $ab$ denotes the $A + a \rightarrow B + b$ reaction channel and subscript $c$ denotes the $B \rightarrow C + c$ decay channel.
$P_{c}$ is the penetrability of the decay channel and the feeding factor, $G_{ab}$, captures the population strength and excitation-energy dependence for the incoming reaction channel (see Ref. \cite{PhysRevC.105.024308} for details).
The total width of the $\mu^\textrm{th}$ level is expressed as a sum over the decay-channel widths
%------------------------------------------------
%%%%	\textbf{R}-matrix Resonance Width
\begin{equation}
  \centering
% 	\Gamma_{1}(E) = \sum\limits_{i} 2\gamma_{i}^{2}P_{i}(E),
	\Gamma_{\mu}(E) = \sum\limits_{c^{\prime}} 2\gamma_{\mu c^{\prime}}^{2}P_{c^{\prime}}(\ell, E),
  \label{eq:FormalTotalWidth}
\end{equation}
%------------------------------------------------
\noindent where $\gamma^{2}$ is the reduced width and $c^{\prime}$ is a summation index over the decay channels.
The penetrability for decay channel $c$, with an orbital angular momentum of the decay, $\ell$, is expressed as
%------------------------------------------------
\begin{equation}
  \centering
	P_{c}(\ell, E) = \frac{ka_{c}}{F_{l}(\eta, ka_{c})^{2} + G_{l}(\eta, ka_{c})^{2}},
  \label{eq:Penetrability}
\end{equation}
%------------------------------------------------
% \noindent where $F_{l}(\eta, ka_{c})$ and $G_{l}(\eta, ka_{c})$ are the regular and irregular Coulomb functions, respectively, $k$ is the wavenumber and $a_{c}$ is the fixed channel radius.
\noindent where $F_{l}(\eta, ka_{c})$ and $G_{l}(\eta, ka_{c})$ are the regular and irregular Coulomb functions, respectively; $k$ is the wavenumber, $a_{c}$ is the fixed channel radius and $\eta$ is the dimensionless Sommerfeld parameter \cite{BreitG1967}.

In the case of an isolated resonance, the corresponding lineshape corresponds to a single-level approximation \cite{RevModPhys.30.257} of the form
%%%%	Lineshape
\begin{equation}
   \centering
    N_{ab,c}(E) = \frac{G_{ab} \, \Gamma_{c} }{\left(E - E_{r} - \Delta\right)^{2} + \frac{1}{4}\Gamma^{2}},
    % N_{ab,c}(E) = \frac{G_{ab} \, \Gamma_{c} }{\left(E - E_{r} - \Delta\right)^{2} + \Gamma^{2}/4},
   \label{eq:SinglelevelApproximation}
\end{equation}
\noindent where $E_{r}$ is the resonance energy and $\Delta \equiv \Delta_{11}$ is expressed as a sum over the decay channels, and
%%%%	Delta
\begin{equation}
  \centering
	\Delta_{\lambda \mu} = \sum\limits_{c^{\prime}} - (S_{c^{\prime}} - B_{c^{\prime}}) \gamma_{\lambda c^{\prime}}\gamma_{\mu c^{\prime}},
  \label{eq:Delta1}
\end{equation}

\noindent where $S_{c}$ and $B_{c}$ are the shift factors and boundary condition parameters, respectively.
For this work, the ``natural'' boundary condition, $B_{c} = S_{c}(E_{r})$, was employed.
The shift factors are expressed as

\begin{equation}
  \centering
	S_{l}(E) = \frac{ka_{c}\left[ F_{l}(\eta, ka_{c})F_{l}^{\prime}(\eta, ka_{c}) + G_{l}(\eta, ka_{c})G_{l}^{\prime}(\eta, ka_{c}) \right]}{F_{l}(\eta, ka_{c})^{2} + G_{l}(\eta, ka_{c})^{2}},
  \label{eq:ShiftFactor}
\end{equation}

\noindent where $F_{l}^{\prime}$ and $G_{l}^{\prime}$ are the derivatives of the regular and irregular Coulomb functions, respectively.

As this study is focused on precisely extracting the physical total width of the $3_{1}^{-}$ resonance, it is important to understand the nuances between the various width definitions relevant to \textbf{R}-matrix analyses.
For the case of an isolated resonance, the full width half maximum (FWHM) for the intrinsic lineshape of a resonance is  referred to as the physical (or intrinsic) total width.
The physical total width is model independent and corresponds to the ENSDF definition of the total width.
% The formal total width $\Gamma_{\mu}(E)$ defined in Eq. \ref{eq:FormalTotalWidth} is correlated with, yet distinct from the physical total width.
The formal total width (Eq. \ref{eq:FormalTotalWidth}) is a highly model-dependent \textbf{R}-matrix quantity which is correlated with, yet distinct from the physical total width.
This is predominantly due to the energy dependence of the energy shift ($\Delta$) in Eq. \ref{eq:SinglelevelApproximation}.
% However, for a given formal total width, the physical total width can be well approximated by what is known as the observed total width, expressed as
However, for a given formal total width, the corresponding physical total width can be well approximated by what is known as the observed total width, defined as
% For the case of an isolated resonance, the observed (or physical) intrinsic full width half maximum (FWHM) of a resonance is correlated with, yet distinct from, the formal total width $\Gamma_{\mu}(E)$ defined in Eq. \ref{eq:FormalTotalWidth}.
% The ENSDF definition for the listed total width is the full width at half maximum (FWHM) intensity for a resonance.
% The physical total width is correlated with, yet distinct from, 
% The expression for $\Gamma_{\mu}(E)$ in Eq. \ref{eq:FormalTotalWidth} is known as the formal total width of a resonance.
% An expression for the observed total width can be produced by 
% Under certain conditions, the physical total width of a resonance can be well approximated using the formal total width to yield the 
% \noindent the observed total width is then expressed as
%------------------------------------------------
\begin{equation}
  \centering
% 	\Gamma_{1}(E) = \sum\limits_{i} 2\gamma_{i}^{2}P_{i}(E),
	\Gamma_{\textrm{obs}, \mu}(E) = \frac{\sum\limits_{c^{\prime}} 2\gamma_{\mu c^{\prime}}^{2}P_{c^{\prime}}(\ell, E)}{1 + \sum\limits_{c^{\prime}} \gamma_{\mu c^{\prime}}^{2} \frac{dS_{c^{\prime}}}{dE} \big|_{E = E_{r}}}.
  \label{eq:ObservedTotalWidth}
\end{equation}
%------------------------------------------------
This approximation is valid under the aforementioned natural boundary condition, and the approximation of the shift factor to be linear in the vicinity of the resonance energy (known as the Thomas approximation \cite{RevModPhys.30.257}) as:
%------------------------------------------------
\begin{equation}
  \centering
	\Delta_{\mu}(E) \approx \Delta_{\mu}(E_{r}) + (E_{r} - E)\sum\limits_{c^{\prime}} \gamma_{\mu c^{\prime}}^{2} \frac{dS_{c^{\prime}}}{dE} \big|_{E = E_{r}}.
	% \Delta(E) \approx \Delta(E_{r}) + (E_{r} - E)\sum\limits_{c^{\prime}} \gamma_{\mu c^{\prime}}^{2} \frac{dS_{c^{\prime}}}{dE} \big|_{E = E_{r}},
  \label{eq:ThomasApproximation}
\end{equation}
%------------------------------------------------
As long as the shift factor and penetrability vary slowly over the resonance range, the observed total width well approximates the physical total width for a Breit-Wigner resonance.
The formal total width is therefore a fundamentally different quantity from the observed total width and the physical total width (FWHM) of an isolated resonance; the latter two widths converging when the Thomas approximation is accurate.
This Thomas approximation is poor for highly clustered resonances located near particle threshold (an example is given in Section \ref{subsubsec:Isolated analysis of 12C_p_p_12C data}).
In such cases, the corresponding shift factors are significantly non-linear across the range of the resonance and the reduced widths are large, which further amplifies the effect of the $\Delta_{\lambda \mu}$ parameter (see Eq. \ref{eq:Delta1}).
In general, the formal total width should therefore not be compared with the physical width observed for a resonance (although these quantities can be very similar under certain conditions).
This feature for this parameterization of \textbf{R}-matrix theory has been detailed in the seminal work of Lane and Thomas \cite{RevModPhys.30.257} as well as in more recent studies such as Refs. \cite{RevModPhys.89.035007, PhysRevC.102.034328}.
% It is therefore standard practice that the resonance energies and widths stored in the ENSDF correspond only to the physical properties of a resonance, i.e. the observed 
It is therefore standard practice for only the physical resonance properties (i.e. the observed parameters) to be stored in evaluated nuclear data libraries such as the ENSDF.
This ensures the portability of results between different analyses (e.g. for ENSDF evaluations), which often employ different channel radii to produce formal parameters which cannot be directly compared.
This is particularly appropriate for \textbf{R}-matrix parameterizations, for which there is no particular ``correct'' channel radius, rather a range of channel radii (related to the physical particles sizes of the partition) which enable data to be well parameterized.
In the context of \textbf{R}-matrix cross-section calculations for nuclear astrophysics, it is standard to treat the ENSDF parameters as observed parameters.
These evaluated quantities can then be transformed to their formal counterparts for any particular channel radius for subsequent calculations.
For completeness, the \textbf{R}-matrix formalism discussed thus far, that requires the choice of arbitrary boundary condition and channel-radius parameters, corresponds to the Wigner-Eisenbud parameterization of \textbf{R}-matrix theory.
Alternative \textbf{R}-matrix parameterisations have been proposed by Brune \cite{PhysRevC.66.044611}, as well as Ducru and Sobes \cite{PhysRevC.105.024601}, which elegantly mitigate ambiguities stemming from these arbitrary parameters.
% Several alternative \textbf{R}-matrix parameterisations have been proposed which elegantly mitigate ambiguities stemming from these arbitrary parameters \cite{PhysRevC.66.044611, PhysRevC.105.024601}.
However, the Wigner-Eisenbud parameterization applied in this work is still widely employed.

%================================================
\subsection{\label{subsec:The primary analysis of inclusive spectra}The primary analysis of inclusive spectra}

\begin{figure}[h!]
\includegraphics[width=\columnwidth]{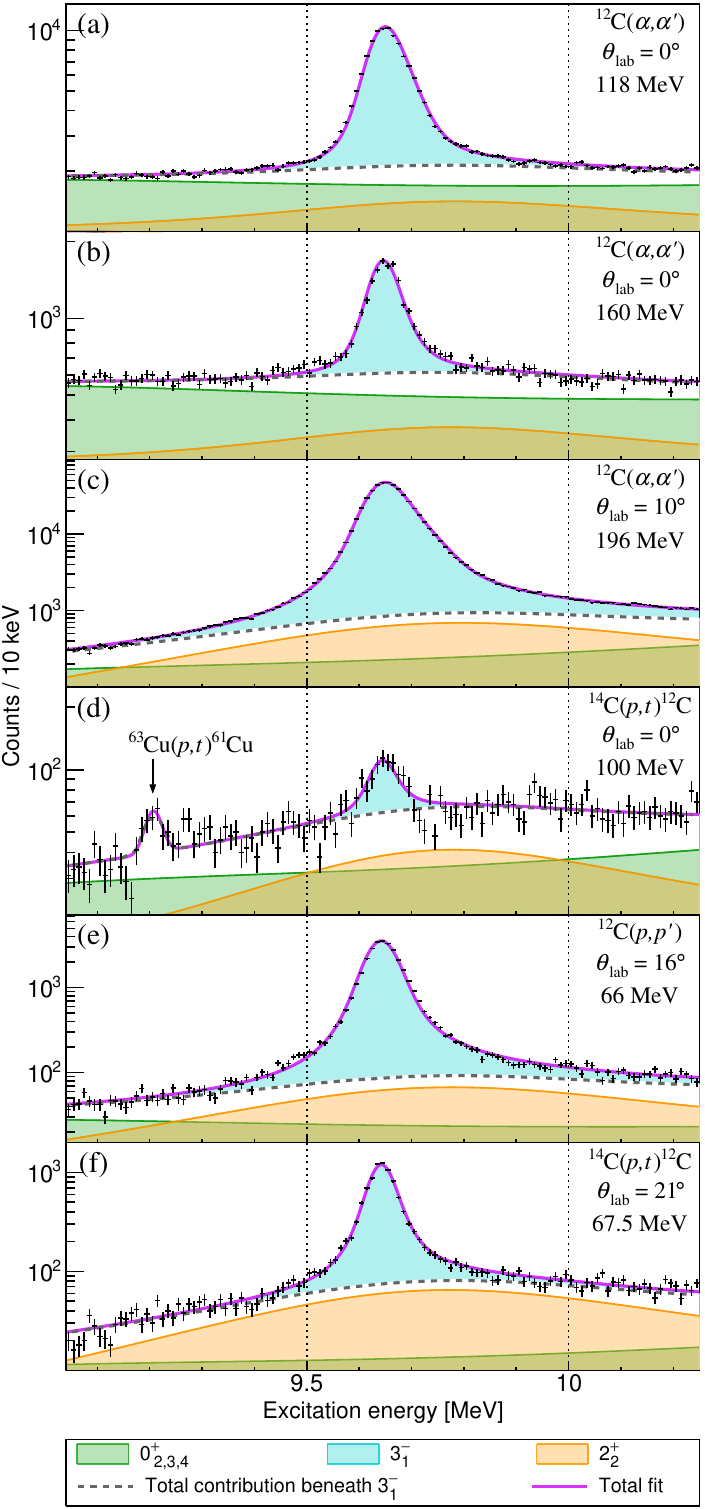}% Here is how to import EPS art
\caption{ \label{fig:AssembleFittedSpectra_GlobalAnalyses_3minus_letter_cropped}
(Color online) Decomposition of the optimal fit for the inclusive excitation-energy spectra.
% and $0_{3}^{+}$ states constructively interfere in the $\alpha_{0}$ channel (see Fig. \ref{fig:AssembledFittedSpectra_MX_PS1_truncated_letter_overpic}).
On each spectrum, the $3_{1}^{-}$ resonance is superimposed on the total contribution from all other resonances and the instrumental background.
For each spectrum, the reaction, measurement angle and beam energy are indicated in the top right of the corresponding panel.
On panel (d), the contaminant state from $\mathrm{^{61}Cu}$ ($E_{x} = 4.756$ MeV) is indicated; see Ref. \cite{PhysRevC.105.024308} for details of the contaminants.
% Panels (a) - (e) correspond to the inclusive ejectile yields for the employed $\mathrm{^{12}C}(\alpha, \alpha^{\prime})\mathrm{^{12}C}$ and $\mathrm{^{14}C}(p, t)\mathrm{^{12}C}$ reactions at various measurement angles and beam energies.
}
\end{figure}

In this work, a simultaneous, self-consistent analysis was performed on the data corresponding to Table \ref{tab:SummaryOfExperimentalParameters}.
The two main contributions to the broad underlying background are the $2_{2}^{+}$ rotational excitation of the Hoyle state as well as an intricate monopole strength which consists of the ghost of the $0_{2}^{+}$ Hoyle state as well as the $0_{3}^{+}$ and $0_{4}^{+}$ resonances (see Ref. \cite{LI2022136928, PhysRevC.105.024308} for a recent experimental investigation).
The $0_{3}^{+}$ resonance at $E_{x}\approx9$ MeV has been suggested to correspond to the breathing-mode excitation of the Hoyle state \cite{PhysRevC.71.021301, CKurokawa, 10.1093/ptep/ptt048, PhysRevC.94.044319, PhysRevC.107.044304, S_Shihang_2023}, with the higher-energy $0_{4}^{+}$ excitation mode at $E_{x} \approx 11$ MeV corresponding to a predicted bent-arm $3\alpha$ structure \cite{10.1143/PTP.57.1262, PhysRevLett.81.5291, NEFF2004357, PhysRevC.94.044319, RevModPhys.90.035004}.
Two versions of the fits were also explored: the first implements the standard prescriptions for the penetrability and shift factor (Eqs. \ref{eq:Penetrability} and \ref{eq:ShiftFactor}) which assume the $\mathrm{^{8}Be}$ daughter states to be infinitesimally narrow.
The second accounts for the intrinsic widths of the $\mathrm{^{8}Be}$ states (see Ref. \cite{PhysRevC.105.024308} for details).
To investigate the channel-radius dependence for the $3_{1}^{-}$ resonance, a range of channel radii from $a_{c} = 4$ to 11 fm was explored in 0.1 fm increments.
% NOTE, define Wigner limit, compare width to calculation, K=3 band head.
In these fits, the Wigner limit was applied to the $\alpha_{1}$ decay channel of the $3_{1}^{-}$ resonance as the corresponding branching ratio is known to be extremely small \cite{PhysRevC.76.034320, PhysRevC.85.037603, PhysRevC.86.064306}.
The optimal fit results from the simultaneous analysis of all the data corresponding to Table \ref{tab:SummaryOfExperimentalParameters} are presented in Fig. \ref{fig:AssembleFittedSpectra_GlobalAnalyses_3minus_letter_cropped} and Table \ref{tab:FitResults_RMatrixAnalyses}, yielding $\Gamma(E_{r}) = 46(2)$ keV with $\Gamma_{\textrm{obs}}(E_{r}) = 38(2)$ keV for the $3_{1}^{-}$ resonance.
% To investigate this significant difference, a meta-analysis of the studies (Refs. \cite{PhysRevC.87.057307, PhysRevC.86.064306, PhysRev.125.992}) considered in the current ENSDF average is provided in the following subsections.

% This meta-analysis assesses the results from Refs. \cite{PhysRevC.87.057307, PhysRevC.86.064306, PhysRev.125.992}, but not the result of \cite{PhysRev.104.1059}

%========================================================================================
\begin{table*}
\caption{\label{tab:FitResults_RMatrixAnalyses}
Optimized fit results for the global analysis of all considered data (see Fig. \ref{fig:AssembleFittedSpectra_GlobalAnalyses_3minus_letter_cropped}).
}
\begin{ruledtabular}
\begin{tabular}{cccccccccccc}
%---------
% & & & \multicolumn{3}{c}{$\alpha_{0}$} & \multicolumn{3}{c}{$\alpha_{1}$}\\ 
% [0.25ex] \cline{4-6} \cline{7-9} \\ [-1.00ex]
 & & & \multicolumn{3}{c}{$\alpha_{0}$ decay} & \multicolumn{3}{c}{$\alpha_{1}$ decay} & \multicolumn{3}{c}{Total width}\\ 
[0.25ex] \cline{4-6} \cline{7-9} \cline{10-12} \\ [-1.00ex]
% \multicolumn{1}{c}{Analysed} &
% \multicolumn{1}{c}{$P_{c}/\mathcal{P}_{c}$\footnotemark[1]} &
\multicolumn{1}{c}{$^{8}\textrm{Be}$\footnotemark[1]} &
% \multicolumn{1}{c}{$^{8}\textrm{Be}$ state} &
\multicolumn{1}{c}{\hphantom{\footnotemark[2]}AIC\footnotemark[2]} &
\multicolumn{1}{c}{$E_{r}$} &
\multicolumn{1}{c}{$a_{\alpha_{0}}$} &
% \multicolumn{1}{c}{$a$} &
\multicolumn{1}{c}{$\theta_{\alpha_{0}}^{2}$} &
\multicolumn{1}{c}{$\Gamma_{\alpha_{0}}(E_{r})$} &
\multicolumn{1}{c}{$a_{\alpha_{1}}$} &
% \multicolumn{1}{c}{$a$} &
\multicolumn{1}{c}{\hphantom{\footnotemark[3]}$\theta_{\alpha_{1}}^{2}$\footnotemark[3]} &
\multicolumn{1}{c}{$\Gamma_{\alpha_{1}}(E_{r})$} &
\multicolumn{1}{c}{$\Gamma(E_{r})$} &
\multicolumn{1}{c}{$\Gamma_{\textrm{obs}}(E_{r})$} &
\multicolumn{1}{c}{$\Gamma_{\textrm{FWHM}}$} 
\\ [0.25ex]
%---------
% \multicolumn{1}{c}{data} &
\multicolumn{1}{c}{widths} &
% \multicolumn{1}{c}{} &
\multicolumn{1}{c}{} &
% \multicolumn{1}{c}{daughter state} & 
\multicolumn{1}{c}{{[}MeV{]}} &
\multicolumn{1}{c}{{[}fm{]}} &
\multicolumn{1}{c}{} &
\multicolumn{1}{c}{{[}keV{]}} &
\multicolumn{1}{c}{{[}fm{]}} &
\multicolumn{1}{c}{} &
\multicolumn{1}{c}{{[}keV{]}} &
\multicolumn{1}{c}{{[}keV{]}} &
\multicolumn{1}{c}{{[}keV{]}} &
\multicolumn{1}{c}{{[}keV{]}} 
\vspace{2pt} \\
\hline \\ [-2.3ex]
%---------
% $\mathrm{^{12}C}$($p,p^\prime$)$\mathrm{^{12}C}$ &discrete& 1874   & 9.644(2) & 4.9 & 0.260(8) & 45(2) & 4.9 & \hphantom{\footnotemark[2]0}1.0(7)\footnotemark[2]\hphantom{0} & 0(2) & 45(2) & 38(2) & 39(3) \\ [0.25ex] \\ [-2.0ex]
% &finite& 1884   & 9.644(2) & 6.2 & 0.100(2) & 42(2) & 6.2 & \hphantom{0}0.00(8) & 0(2) & 42(2) & 40(2) & 40(2) \\ [0.25ex] \hline \\ [-2.0ex]
% $\mathrm{^{14}C}(p, t)\mathrm{^{12}C}$ &discrete& 1208   & 9.641(2) & 5.2 & \hphantom{0}0.19(1)\hphantom{0} & 42(3) & 5.2 & \hphantom{0}1.0(10) & 0(2) & 42(3) & 37(2) & 37(2) \\ [0.25ex] \\ [-2.0ex]
% % & 1248   & 9.641(2) & 7.1 & 0.059(3) & 37(2) & 7.1 & 0.0(9) & 0(28) & 37(2) & 36(2) & 36(2) \\ [0.25ex] \hline \\ [-2.0ex]
% &finite& 1248   & 9.641(2) & 7.1 & 0.059(3) & 37(2) & 7.1 & \hphantom{0}0.0(9)\hphantom{0} & $\approx$0\footnotemark[2] & 37(2) & 36(2) & 36(2) \\ [0.25ex] \hline \\ [-2.0ex]
Discrete& 15088   & 9.641(2) & 4.8 & 0.291(2) & 46(2) & 4.8 & \hphantom{0}0.0(1)\hphantom{0} & 0(2) & 46(2) & \hphantom{\footnotemark[4]}38(2)\footnotemark[4] & 38(2) \\ [0.25ex] \\ [-2.0ex]
Finite& 15179   & 9.641(2) & 4.8 & 0.316(3) & 47(2) & 4.8 & \hphantom{0}0.00(7) & 0(2) & 47(2) & 39(2) & 39(2) \\ [-0.3ex]
%---------
% \\ [0.25ex] \hline \\ [-2.0ex]
% \\ [-0.3ex]
\end{tabular}
\end{ruledtabular}
\footnotetext[1]{
This indicates the prescription employed for the penetrability and shift factor: ``Discrete'' is the standard prescription which assumes the $\mathrm{^{8}Be}$ states to be infinitely narrow.
``Finite'' indicates that the intrinsic widths of the $\mathrm{^{8}Be}$ states were accounted for \cite{PhysRevC.105.024308}.
% $P_{c}$ indicates that the standard prescriptions for the penetrability and shift factor are employed, which assumed the $\mathrm{^{8}Be}$ daughter states to be infinitely narrow.
% $\mathcal{P}_{c}$ indicates that the penetrability and shift factor account for the finite widths of the $\mathrm{^{8}Be}$ $2_{1}^{+}$ daughter state.
}
\footnotetext[2]{
The AIC estimator is used to determine the best quality model as maximum likelihood estimation is employed to fit the data. 
Minimum chi-square estimation is not used as there are several regions in the spectra which have very low counts ($<10$); Ref. \cite{PhysRevC.105.024308} presents the full fit ranges for a subset of the considered data.
}
\footnotetext[3]{
The Wigner limit is applied for the $\alpha_{1}$ decay channel; see text for details.
}
% \footnotetext[3]{
% In this particular case, the associated uncertainty is ignored for the total width as this somewhat spurious error is highly correlated with other parameters; a consequence of accounting for the finite width of the broad $\mathrm{^{8}Be}$ $2_{1}^{+}$ state.
% }
\footnotetext[4]{
Corresponds to the optimal fit for the $3_{1}^{-}$ resonance in this work.
}
\end{table*}
%------------------------------------------------

%------------------------------------------------
%------------------------------------------------

%================================================================================================
\subsection{\label{subsec:Assessing the $3_{1}^{-}$ total width from Kokalova et al.}Assessing the $3_{1}^{-}$ total width from Ref. \cite{PhysRevC.87.057307}}

To assess the $3_{1}^{-}$ total width reported by in Ref. \cite{PhysRevC.87.057307}, an isolated analysis was performed on the subset of the $\mathrm{^{12}C}(p, p^{\prime})\mathrm{^{12}C}$ data from Ref. \cite{PhysRevC.87.057307}.
These $\mathrm{^{12}C}(p, p^{\prime})\mathrm{^{12}C}$ data are the same that were included in the primary analysis of this work (see Section \ref{subsec:The primary analysis of inclusive spectra} and Table \ref{tab:SummaryOfExperimentalParameters}).
This analysis was repeated for a $\mathrm{^{14}C}(p, t)\mathrm{^{12}C}$ excitation-energy spectrum with better resolution to confirm whether the same observed systematic trends are independent of the data (see Section \ref{subsubsec:Isolated analysis of the 14C_p_t_12C (theta_lab = 21 deg) data}).

%================================================================================================
\subsubsection{\label{subsubsec:Isolated analysis of 12C_p_p_12C data}Isolated analysis of $\mathrm{^{12}C}(p, p^{\prime})\mathrm{^{12}C}$ data}

\begin{figure}[hbtp]
\includegraphics[width=\columnwidth]{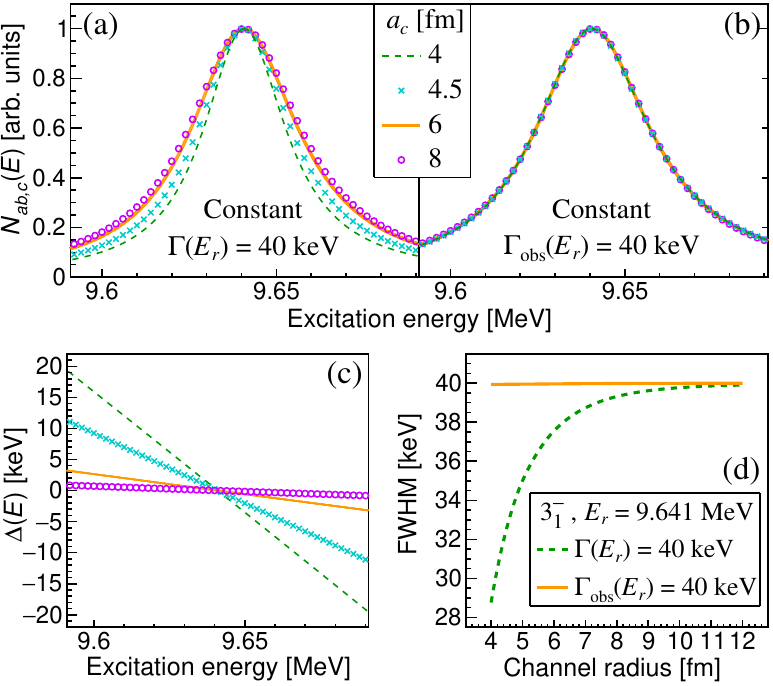}% Here is how to import EPS art
\caption{ \label{fig:Comparison_ChannelRadiusDependence_cropped_9641keV} 
(Color online) The channel-radius dependence of the $3_{1}^{-}$ resonance for the intrinsic lineshape with (a) constant $\Gamma(E_{r}) = 40$ keV, (b) constant $\Gamma_{\textrm{obs}}(E_{r}) = 40$ keV, (c) the energy shift and (d) the FWHM of the intrinsic lineshape.
}
\end{figure}

Ref. \cite{PhysRevC.87.057307} reported only the formal total width for the $3_{1}^{-}$ resonance, which was incorrectly compared to the physical total width of the state reported in other works currently employed in the ENSDF average \cite{PhysRev.104.1059, PhysRev.125.992, PhysRevC.86.064306}.
As previously mentioned in Section \ref{subsec:RMatrixFormalism}, the formal total width is a model-dependent quantity which is distinct from the physical total width (FWHM) of a resonance.
Consequently, this misstated formal total width of $\Gamma = 48(2)$ keV reported by Ref. \cite{PhysRevC.87.057307} was mistakenly considered in the ENSDF average for the physical total width (FWHM) of the $3_{1}^{-}$ resonance \cite{KELLEY201771}.
The current ENSDF average is thus invalid for determining the observed total radiative width for the $3_{1}^{-}$ resonance.
Determining the observed total radiative width requires knowledge on the observed total width, in addition to the $E1$ ($3_{1}^{-} \rightarrow 2_{1}^{+}$) $\gamma$-decay branching ratio and the physical $E3$ ($3_{1}^{-} \rightarrow 0_{\textrm{g.s.}}^{+}$) $\gamma$-decay width of 0.31(4) meV \cite{CRANNELL1967152} (see Ref. \cite{TSUMURA2021136283} for a more detailed discussion).
% To clarify why this misstated result is particularly problematic for the $3_{1}^{-}$ state in $\mathrm{^{12}C}$, an analysis is given in Fig. \ref{fig:Comparison_ChannelRadiusDependence_cropped_9641keV}.
% Panels (a) and (b) of Fig. \ref{fig:Comparison_ChannelRadiusDependence_cropped_9641keV}, present the channel-radius dependence of the $3_{1}^{-}$ intrinsic lineshapes with the formal and observed total widths being kept constant at 40 keV, respectively.
To clarify why this misstated result is particularly problematic for the $3_{1}^{-}$ state in $\mathrm{^{12}C}$, an analysis is given: panels (a) and (b) of Fig. \ref{fig:Comparison_ChannelRadiusDependence_cropped_9641keV}, present the channel-radius dependence of the $3_{1}^{-}$ intrinsic lineshapes with the formal and observed total widths being kept constant at 40 keV, respectively.
%------------------------------------------------
%------------------------------------------------
The clear channel-radius dependence of the formal width is apparent, in contrast to the observed total width which is weakly dependent of the channel radius.
% The channel-radius independence of the observed width is apparent, in contrast to that of the formal width.
This effect is predominantly due the channel-radius dependence of the shift function near the $\alpha_{0}$ decay threshold.
In general, the shift function at a particular energy becomes more constant towards larger channel radii (equivalently for the energy shift, $\Delta$) and thus, the formal total width converges with the observed total width at large channel radii, as shown in panels (c) and (d) of Fig. \ref{fig:Comparison_ChannelRadiusDependence_cropped_9641keV}.

\begin{figure}[hbtp]
\includegraphics[width=\columnwidth]{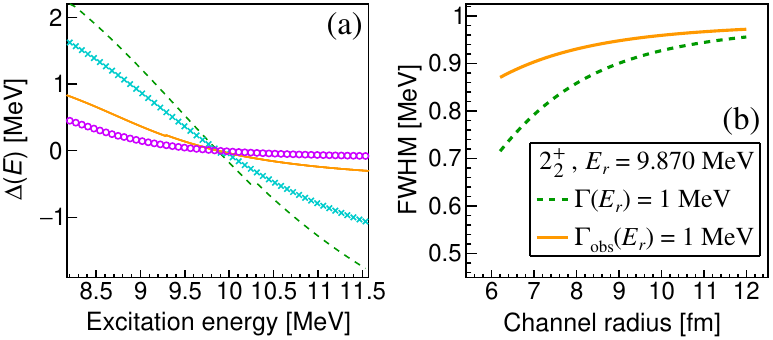}% Here is how to import EPS art
\caption{ \label{fig:Comparison_ChannelRadiusDependence_cropped_9870keV} 
(Color online) The channel-radius dependence of the $2_{2}^{+}$ resonance for (a) the energy shift and (b) the numerically determined FWHM of the intrinsic lineshape.
}
\end{figure}

For completeness, a counter example for the channel-radius independence of the observed total width (defined in Eq. \ref{eq:ObservedTotalWidth}) is provided by the broad $2_{2}^{+}$ rotational excitation of Hoyle state which underlies the $3_{1}^{-}$ resonance, see Fig. \ref{fig:Comparison_ChannelRadiusDependence_cropped_9870keV}.
%------------------------------------------------
%------------------------------------------------
Over the broad range of the $2_{2}^{+}$ resonance, which has an experimentally observed FWHM of $\approx1$ MeV, the energy shift is observed to be non-linear.
Over the range of explored channel radii shown in Fig. \ref{fig:Comparison_ChannelRadiusDependence_cropped_9870keV}, the Thomas approximation is inappropriate for the $2_{2}^{+}$ resonance and thus, $\Gamma_{\textrm{obs}}(E_{r})$ is no longer an accurate measure for the intrinsic FWHM of the resonance.
This general effect manifests strongly for resonances which are located near a particle threshold and exhibit significant clustering (i.e., large reduced widths).
% For such cases, we encourage the reporting of not only the formal and observed total widths, but the numerically calculated FWHMs of the intrinic lineshapes.
For such cases, we thus encourage the reporting of not only the formal and observed width, but also the FWHM ($\Gamma_{\textrm{FWHM}}$) to facilitate model independence in comparisons to data and between analyses.
For \textbf{R}-matrix derived lineshapes, the FWHMs reported in this work are numerically determined for the intrinsic lineshape.
For $\Gamma_{\textrm{FWHM}}$ to be consistent between different direct-reaction datasets, $\Gamma_{\textrm{FWHM}}$ is determined for intrinsic lineshapes without feeding factors.
% In this work, $\Gamma_{\textrm{FWHM}}$ is determined for intrinsic lineshapes without feeding factors to enable consistent comparisons between different direct-reaction datasets.

\begin{figure}[hbtp]
\includegraphics[width=\columnwidth]{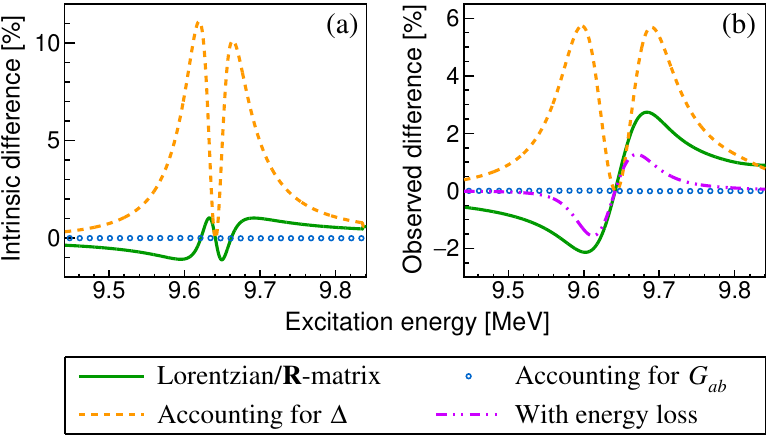}% Here is how to import EPS art
\caption{ \label{fig:Comparison_AnalysisApproximations_new_cropped} 
(Color online) The differences in the (a) intrinsic and (b) experimentally observed lineshapes caused by the considered approximations (relative to the peak maximum).
% For the \textbf{R}-matrix lineshapes, an observed total width of $\Gamma_{\textrm{obs}}(E_{r}) = 40$ keV at $E_{r} = 9.641$ MeV was employed.
% For the standard Lorentzian, an energy-independent width of $\Gamma = 40$ keV was implemented.
% A constant observed total width of $\Gamma_{\textrm{obs}}(E_{r}) = 40$ keV at $E_{r} = 9.641$ MeV was employed for all \textbf{R}-matrix lineshapes; for the standard Lorentzian, an energy-independent width of $\Gamma = 40$ keV was implemented.
The observed lineshapes were produced by convoluting the intrinsic lineshapes with the experimental response observed in the $\mathrm{^{12}C}$($p,p^\prime$)$\mathrm{^{12}C}$ data analyzed in this work.
}
\end{figure}

In this work, the formal total width of the $3_{1}^{-}$ resonance of $\Gamma(E_{r}) = 48(2)$ keV reported by Ref. \cite{PhysRevC.87.057307} has been converted to the observed total width of $\Gamma_{\textrm{obs}}(E_{r}) = 39(4)$ keV which is appropriate for the ENSDF average. This observed total width matches the numerically determined FWHM of $\Gamma_{\textrm{FWHM}} = 39(4)$ keV (see Table \ref{tab:DataUsedForNNDCEvaluationOfWidth}).
The overall uncertainties for these converted values include previously unaccounted-for systematic errors stemming from the approximations employed in Ref. \cite{PhysRevC.87.057307}.
% A number of analysis approximations were employed in Ref. \cite{PhysRevC.87.057307} and to check whether this corrected observed total width should be employed in future ENSDF evaluations, 
To estimate these unaccounted-for systematic errors, a comprehensive, isolated analysis was also performed on the subset of the data employed in Ref. \cite{PhysRevC.87.057307}.
This subset exhibits a better experimental resolution of 48(1) keV (FWHM) and fewer experimental artefacts than in Ref. \cite{PhysRevC.87.057307}.
The study of Ref. \cite{PhysRevC.87.057307} parameterized the intrinsic lineshape for the $3_{1}^{-}$ resonance with that of a single, isolated level (Eq. \ref{eq:SinglelevelApproximation}).
This is a reasonable approximation given the relatively narrow width of $\approx 40$ keV with respect to the next closest $3^{-}$ resonance at $E_{x} = 18350(50)$ keV.
Furthermore, the single (decay) channel approximation employed by Ref. \cite{PhysRevC.87.057307} is appropriate as the $3_{1}^{-}$ resonance is understood to decay almost exclusively through the $\alpha_{0}$ ($\ell=3$/$f$-wave) decay channel to the $0^{+}$ ground state of $\mathrm{^{8}Be}$ \cite{PhysRevC.76.034320, PhysRevC.85.037603, PhysRevC.86.064306}.
The \textbf{R}-matrix-derived lineshape employed in Ref. \cite{PhysRevC.87.057307} is functionally the same as Eq. \ref{eq:SinglelevelApproximation}, with Ref. \cite{PhysRevC.87.057307} having employed a degenerate sign difference in the definition of both $\Delta$ and the first term of the denominator in Eq. \ref{eq:SinglelevelApproximation}.
Perhaps the most critical approximation made by Ref. \cite{PhysRevC.87.057307} concerns the description of the broad, background components beneath the $3_{1}^{-}$ resonance with a second-order polynomial and a $2^{+}$ resonance situated at $E_{x} \approx 9.6$ MeV with $\Gamma \approx 600$ keV.
The effect of target-related energy loss for the ejectile is also ignored. 
Finally, the feeding factor for the direct populating channel ($G_{ab}$) was also implicitly approximated to be constant.
Ref. \cite{PhysRevC.87.057307} also investigated whether the intrinsic $3_{1}^{-}$ lineshape can be well parameterized with a standard (symmetric) Lorentzian in comparison to the \textbf{R}-matrix lineshape given by Eq. \ref{eq:SinglelevelApproximation}.
In this investigation, we also consider the somewhat-common approximation of employing a pseudo \textbf{R}-matrix lineshape where the energy shift ($\Delta$ in Eq. \ref{eq:SinglelevelApproximation}) is neglected.
A systematic sensitivity study of these approximations was performed for both the intrinsic and experimentally observed lineshapes of the $3_{1}^{-}$ resonance, see Fig. \ref{fig:Comparison_AnalysisApproximations_new_cropped}.
A Gaussian experimental resolution of $\sigma = 20$ keV was employed, which is similar to the $\sigma = 23(1)$ keV resolution reported in Ref. \cite{PhysRevC.87.057307}.
%------------------------------------------------
%------------------------------------------------
%------------------------------------------------
\begin{table*}
\caption{\label{tab:FitResults_IsolatedAnalyses_PR146_16deg}
Summary of analysis configurations and results for the $\mathrm{^{12}C}$($p,p^\prime$)$\mathrm{^{12}C}$ reaction at $E_{\beam} = 66$~MeV.
$\goodchi^{2}_{\textrm{red}}$ is the reduced chi-squared statistic.
The Landau energy-loss parameters were fitted to the intrinsically narrow Hoyle state in the subset of $\mathrm{^{12}C}$($p,p^\prime$)$\mathrm{^{12}C}$ data from Ref. \cite{PhysRevC.87.057307} analyzed in this work.
The channel radius for all \textbf{R}-matrix derived lineshapes is the same as that employed by Ref. \cite{PhysRevC.87.057307}: $a_{c} = 1.3(4^{1/3} + 8^{1/3}) \approx 4.7$ fm.
The reported errors contain the fitting errors and a minimum focal-plane detection error of 1 keV, added in quadrature, with no other systematic errors included (e.g., channel-radius dependence).
}
\begin{ruledtabular}
\begin{tabular}{clccccccccc}
%---------
& & & & & & & \multicolumn{4}{c}{$3_{1}^{-}$}
\\ [0.25ex] \cline{8-11} \\ [-1.00ex]
\multicolumn{1}{c}{Case} &
\multicolumn{1}{c}{Background} &
% \multicolumn{2}{c}{Convolution} &
\multicolumn{1}{c}{Intrinsic} &
% \multicolumn{1}{c}{Target-related} &
% \multicolumn{1}{c}{$\Delta_{c} \neq 0$} &
\multicolumn{1}{c}{Energy} &
\multicolumn{1}{c}{Energy} &
% \multicolumn{1}{c}{Assume} &
% \multicolumn{1}{c}{Approx.} &
% \multicolumn{1}{c}{Account} &
\multicolumn{1}{c}{Feeding} &
\multicolumn{1}{c}{$\goodchi^{2}_{\textrm{red}}$} &
% \multicolumn{1}{c}{$\Gamma(E_{r})$ / $\Gamma_{\textrm{obs}}(E_{r})$ / $\textrm{FWHM}$} & 
% \multicolumn{1}{c}{$\Gamma_{\FWHM}$, $2_{2}^{+}$} &
\multicolumn{1}{c}{$E_{r}$} &
\multicolumn{1}{c}{$\Gamma(E_{r})$} &
\multicolumn{1}{c}{$\Gamma_{\textrm{obs}}(E_{r})$} &
\multicolumn{1}{c}{$\Gamma_{\textrm{FWHM}}$} 
% \multicolumn{1}{c}{$\Gamma(E_{r})/\Gamma_{\textrm{obs}}(E_{r})/\textrm{FWHM}$} 
\\ [0.25ex]
%---------
\multicolumn{1}{c}{} &
\multicolumn{1}{c}{} &
\multicolumn{1}{c}{lineshape, $3_{1}^{-}$} &
\multicolumn{1}{c}{shift, $\Delta$} &
\multicolumn{1}{c}{loss} &
\multicolumn{1}{c}{factors} & 
\multicolumn{1}{c}{} & 
\multicolumn{1}{c}{{[}MeV{]}} &
\multicolumn{1}{c}{{[}keV{]}} &
\multicolumn{1}{c}{{[}keV{]}} &
\multicolumn{1}{c}{{[}keV{]}} 
\vspace{2pt} \\
%---------
\hline \\ [-2.3ex]
{(1)}   & Quadratic    & \textbf{R}-matrix   & \checkmark      &               &                & 1.669   & 9.645(1)& 54(1)& 43(1)& 43(1) \\ [0.25ex] \hline \\ [-2.0ex]
%----------------------------
{(2)}   & Quadratic + Gaussian& Lorentzian   &                  &               &                & 7.209   & 9.646(1)& 36(1)&  &36(1) \\ [0.25ex]  \\ [-2.0ex]
{(3)}   & Quadratic + Gaussian& \textbf{R}-matrix   &                  &               &                & 1.310   & 9.645(1)& 41(1) & 41(1) & 41(1) \\ [0.25ex]  \\ [-2.0ex]
{(4)}   & Quadratic + Gaussian& \textbf{R}-matrix   & \checkmark      &               &                & 1.341   & 9.645(1)& 51(1)& 41(2)& 41(1) \\ [0.25ex]  \\ [-2.0ex]
{(5)}   & Quadratic + Gaussian& \textbf{R}-matrix   & \checkmark      & \checkmark   &                & 1.321   & 9.644(1)& 50(1)& 41(2)& 41(1) \\ [0.25ex]  \\ [-2.0ex]
{(6)}   & Quadratic + Gaussian& \textbf{R}-matrix   & \checkmark      & \checkmark   & \checkmark    & 1.321   & 9.644(1)& 50(1)& 41(2)& 41(1) \\ [0.25ex] \hline \\ [-2.0ex]
%----------------------------
{(7)}   & Quadratic + Lorentzian & \textbf{R}-matrix   & \checkmark      &               &                & 1.358   & 9.645(1)& 51(1)& 41(2)& 41(1) \\ [0.25ex] \hline \\ [-2.0ex]
%----------------------------
{(8)}   & Quadratic + $2_{2}^{+}$ (Eq. \ref{eq:SinglelevelApproximation}) & \textbf{R}-matrix   & \checkmark      &               &                & 1.269   & 9.645(1)& 47(2)& 38(2)& 38(2) \\ [-0.3ex]
%====================================================================
\end{tabular}
\end{ruledtabular}
\end{table*}
%------------------------------------------------
For the \textbf{R}-matrix lineshape of Eq. \ref{eq:SinglelevelApproximation}, an observed total width of $\Gamma_{\textrm{obs}}(E_{r}) = 40$ keV at $E_{r} = 9.641$ MeV was employed.
For the test of a standard (symmetric) Lorentzian, an energy-independent width of $\Gamma = 40$ keV was implemented.
It is observed that the inclusion (or lack) of an energy-dependent feeding factor is negligible. 
This is unsurprising given the typical slowly-varying energy dependence of the feeding factor \cite{PhysRevC.105.024308}, which is well approximated to be constant over the $\approx 40$ keV width of the $3_{1}^{-}$ resonance.
The effects of employing a standard Lorentzian and neglecting the target-related energy loss are more significant given the small statistical errors for the high yields at the peak of the $3_{1}^{-}$ resonance in Ref. \cite{PhysRevC.87.057307} ($\approx 20000$ counts per 5 keV bin).
The $\Delta = 0$ approximation was observed to have the most dramatic effect on both the intrinsic and experimentally observed lineshape for the $3_{1}^{-}$ resonance.
In general, this approximation is particularly poor for resonances near threshold which exhibit a high degree of clustering, and thus large reduced widths and quickly varying energy shifts.
To clarify this, the $\Delta = 0$, pseudo \textbf{R}-matrix approximation to the intrinsic lineshape was not explored in Ref. \cite{PhysRevC.87.057307}, however it is detailed here as a caution for future studies which may seek to apply it.
% Whilst it is clear that several of these approximations may have a significant effect on the observed lineshape, the associated effect on the extracted observed total width for the $3_{1}^{-}$ resonance is dependent on the features of the data.
It is clear that several of these approximations may individually have a significant effect on the observed lineshape.
However, the combined effect from these approximations on the extracted observed total width for the $3_{1}^{-}$ resonance is dependent on the features of the data.
To gauge the systematic errors associated with these various approximations, an exhaustive test for various combinations of these approximations was performed on a subset of the $\mathrm{^{12}C}$($p,p^\prime$)$\mathrm{^{12}C}$ data in Ref. \cite{PhysRevC.87.057307} and the most revealing combinations are presented as different cases in Table \ref{tab:FitResults_IsolatedAnalyses_PR146_16deg}.
The associated fits were performed with $\goodchi^{2}_{\textrm{red}}$ minimization given the limited fit range (which approximately matches that in Ref. \cite{PhysRevC.87.057307}) in which the data is approximately normally distributed.

% for which the uncertainties are approximately normally distributed (in contrast to the global analysis in Table \ref{tab:FitResults_RMatrixAnalyses}).

% Unlike the global analysis in Table \ref{tab:FitResults_RMatrixAnalyses} which fitted a large excitation-energy range with regions of of low counts, the fits in Table \ref{tab:FitResults_IsolatedAnalyses_PR146_16deg} were performed with $\goodchi^{2}_{\textrm{red}}$ minimization as fit range , 

In terms of background systematics, it is observed that the choice of parameterization for the underlying $2_{2}^{+}$ contribution is significant.
First-order and second-order polynomial backgrounds were found to yield almost identical fits, thus the results from the former are omitted from Table \ref{tab:FitResults_IsolatedAnalyses_PR146_16deg}.

\begin{figure}[bp]
\includegraphics[width=\columnwidth]{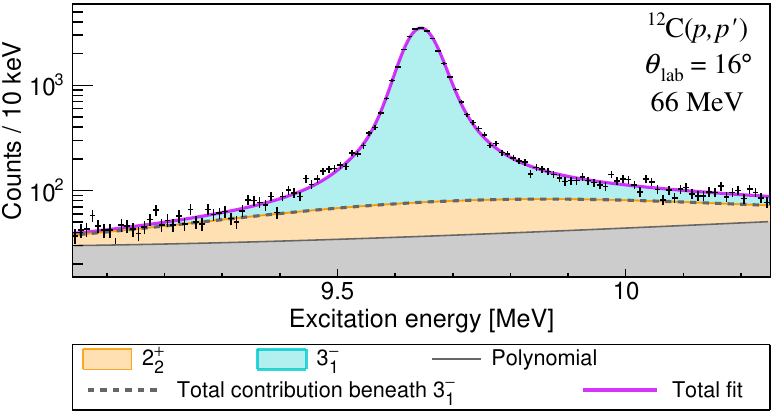}% Here is how to import EPS art
\caption{ \label{fig:AssembleFittedSpectra_IsolatedAnalyses_3minus_letter_only_12C_p_p_12C_cropped}
(Color online)
Decomposition of the fit for case (4) in Table \ref{tab:FitResults_IsolatedAnalyses_PR146_16deg}.
The $3_{1}^{-}$ resonance is superimposed on the total contribution of the $2_{2}^{+}$ resonance (parameterized with a Gaussian lineshape) and a polynomial background.
The reaction, measurement angle and beam energy are indicated.
}
\end{figure}

\begin{itemize}
    
    \item \textbf{Case (1):} The observed total width for the $3_{1}^{-}$ resonance is larger when approximating the underlying $2_{2}^{+}$ resonance with simplified Gaussian and standard Lorentzian lineshapes and larger still when not including a broad component for the $2_{2}^{+}$ resonance at all.
    
    \item \textbf{Case (2):} In terms of the lineshape for the $3_{1}^{-}$ resonance itself, a standard (symmetric) Lorentzian lineshape [case (2)] yields a poor fit and exhibits a significantly smaller $\Gamma_{\textrm{FWHM}}$ than other scenarios.

    \item \textbf{Case (3):} A comparison of case (3) to case (4) cases shows that ignoring the energy shift naturally affects the extracted formal width, however, the FWHM of the $3_{1}^{-}$ resonance is consistent.

    \item \textbf{Case (4):} This case most closely matches the fit methodology employed in Ref. \cite{PhysRevC.87.057307} and the corresponding fit in Fig. \ref{fig:AssembleFittedSpectra_IsolatedAnalyses_3minus_letter_only_12C_p_p_12C_cropped} yielded $\Gamma(E_{r}) = 51(1)$ keV and $\Gamma_{\textrm{obs}}(E_{r}) = 41(2)$ keV, with the formal total width being in reasonable agreement with the corresponding value reported in Ref. \cite{PhysRevC.87.057307} (see Table \ref{tab:DataUsedForNNDCEvaluationOfWidth}).
    In this work, the formal total reported in Ref. \cite{PhysRevC.87.057307} was converted to the appropriate observed total width for the ENSDF.
    By comparing the fit of case (4) to the best fit of case (8), the unaccounted-for systematic error due to the background approximation in Ref. \cite{PhysRevC.87.057307} can be roughly estimated at 3 keV for both $\Gamma_{\textrm{obs}}(E_{r})$ and $\Gamma_{\textrm{FWHM}}$.
    This error has been included for the corrected values of $\Gamma_{\textrm{obs}}(E_{r})= 39(4)$ keV and $\Gamma_{\textrm{FWHM}} = 39(4)$ keV which we recommend for future evaluations which consider the result of Ref. \cite{PhysRevC.87.057307}.

    \item \textbf{Case (5):} The inclusion of the target-related energy loss improves the quality of the fit, however the extracted formal and observed widths are not strongly affected given the energy resolution of the data.

    \item \textbf{Case (6):} The feeding factors have a negligible effect on the fits in general. 
    This is to be expected given the typical slowly-varying energy dependence of the feeding factor \cite{PhysRevC.105.024308}, which is well approximated to be constant over the $\approx 40$ keV width of the $3_{1}^{-}$ resonance.

    \item \textbf{Case (7):} Approximating the underlying $2_{1}^{+}$ resonance as a symmetric Lorentzian yields a poorer fit in comparison to the Gaussian approximation for the $2_{1}^{+}$ resonance in case (4).
    This may be due to the fact that the intrinsic shape of the $2_{1}^{+}$ is asymmetric, with the low-energy tail being strongly suppressed by the penetrability; a feature that the longer tails of a Lorentzian (in comparison to a Gaussian) are less suited to approximate.

    \item \textbf{Case (8):} This provides the best fit of the considered cases, employing the single-level, single-channel approximation of Eq. \ref{eq:SinglelevelApproximation} for both the $3_{1}^{-}$ resonance and the underlying broad $2_{2}^{+}$ resonance, yielding a formal total width of $\Gamma(E_{r}) = 47(2)$ keV, with an observed total width of $\Gamma_{\textrm{obs}}(E_{r}) = 38(2)$ keV.
    To further emphasize the inappropriate nature of comparing the channel-radius dependent formal total width to the observed total width (see Fig. \ref{fig:Comparison_ChannelRadiusDependence_cropped_9641keV}), a fit with case (8) using a channel-radius of 4 fm was performed as opposed to the choice of 4.7 fm (see caption of Table \ref{tab:FitResults_IsolatedAnalyses_PR146_16deg}) in Ref. \cite{PhysRevC.87.057307}, yielding a much larger formal total width of $\Gamma(E_{r}) = 65(4)$ keV, whilst the observed total width of $\Gamma_{\textrm{obs}}(E_{r}) = 40(2)$ keV is in good agreement with the optimal fit in this work which yielded $\Gamma_{\textrm{obs}}(E_{r}) = 38(2)$ keV.
    
\end{itemize}

%------------------------------------------------
%------------------------------------------------

% As the choice of channel radius of $a_{c} = 1.3(4^{1/3} + 8^{1/3}) \approx 4.7$ {[fm]} in Ref. \cite{PhysRevC.87.057307} is completely arbitrary, the systematic error stemming from the channel-radius dependence was estimated by performing fits with case (4) over a range of channel radii from $a_{c} = 4$ to 11 fm in 0.1 fm increments.

\begin{figure}[hbtp]
\includegraphics[width=\columnwidth]{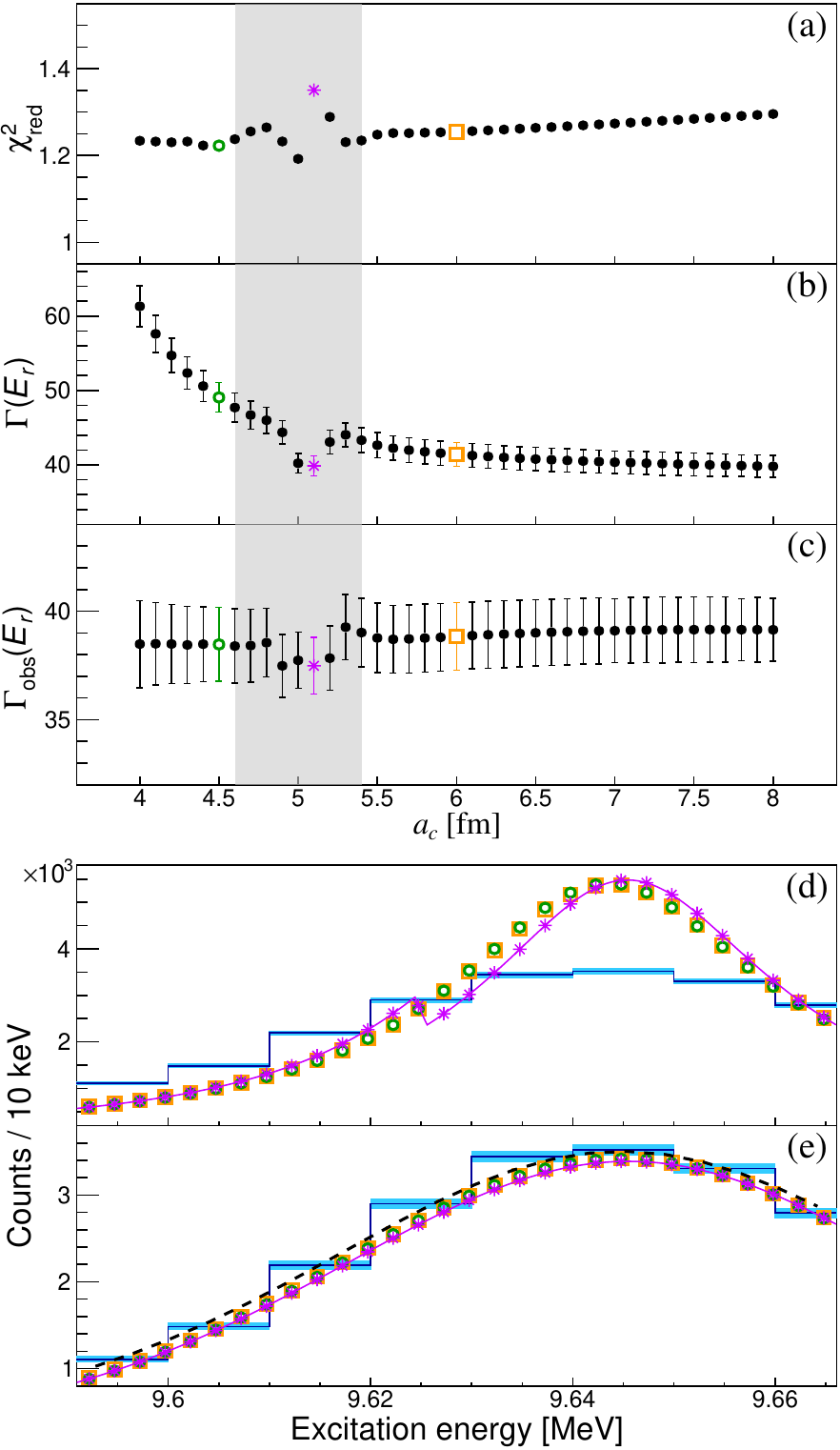}% Here is how to import EPS art
\\
\includegraphics[width=\columnwidth]{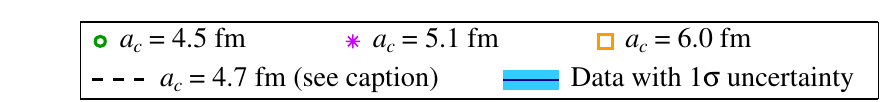}% Here is how to import EPS art
\caption{ \label{fig:Comparison_FitMetricsAndWidths_ChannelRadiusDependence_cropped} 
(Color online) For the $3_{1}^{-}$ resonance: the channel-radius dependence of (a) $\goodchi^{2}_{\textrm{red}}$, (b) $\Gamma(E_{r})$ and (c) $\Gamma_{\textrm{obs}}(E_{r})$.
The range of channel radii which yield significant discontinuities (4.6--5.4 fm) is highlighted in filled gray.
Panels (d) and (e) respectively present the intrinsic and observed lineshapes corresponding to the channel radii of 4.5, 5.1 and 6.0 fm (see legend), corresponding to panels (a)-(c).
The fit results at different channel radii correspond to the fit conditions of case (4) (see Table \ref{fig:Comparison_AnalysisApproximations_new_cropped}), with the exception of $a_{c} = 4.7$ fm in panel (e) which includes the Landau energy loss of case (5).
% The exception is the black dashed line in panel (e) which employs the conditions of case (5) to show the improved fit by including a Landau energy loss in the observed lineshape.
% The exception is the black dashed line in panel (e) which employs the conditions of case (5).
}
\end{figure}

To emphasize the significant channel-radius dependence of the formal total width for the $3_{1}^{-}$ resonance (in contrast to the observed total width), fits were performed with case (4) over a range of channel radii from $a_{c} = 4$ to 8 fm in 0.1 fm increments (see Fig. \ref{fig:Comparison_FitMetricsAndWidths_ChannelRadiusDependence_cropped}). 
%------------------------------------------------
%------------------------------------------------
The optimal channel radius for this isolated fit was determined as $a_{c} = 4.5$ fm, with all explored channel radii providing a similar quality fit.
As previously mentioned, whilst the difference between formal and observed widths can sometimes be quite small, the difference can be significant under certain conditions.
This is precisely the case for the $3_{1}^{-}$ resonance, for which a choice of $a_{c} = 4.0$ is demonstrated to describe the data similarly well in Fig. \ref{fig:Comparison_FitMetricsAndWidths_ChannelRadiusDependence_cropped}(a), would yield a formal width greater than 60 keV that is unrelated to the physical total width of the state.
The observed fluctuations in the reduced chi-squared for some channel radii between $a_{c} = 4.6$--5.4 fm is a consequence of small numerical inaccuracies for the Coulomb functions calculated with the GNU Scientific Library (GSL) in this work \cite{GNU}.
As such, the fits with strongly-deviating channel radii (e.g. $a_{c} = 5.0$--5.2 fm) were disregarded as the numerical inaccuracies for these cases were more significant.
An example of how these numerical inaccuracies affect the intrinsic and experimental lineshapes for the $3_{1}^{-}$ resonance is presented in Fig. \ref{fig:Comparison_FitMetricsAndWidths_ChannelRadiusDependence_cropped}(d) and (e), respectively.
It is observed that the numerical error for $a_{c} = 5.1$ presents as a jagged ``discontinuity'' in the intrinsic lineshape near $E_{x} = 9.625$ MeV.
The scale and location of these numerical inaccuracies are highly dependent on the affected $a_{c}$ values, however, these features are generally more prevalent towards lower energies.
For most cases, these numerical inaccuracies are orders of magnitude below the statistical errors of the considered data and therefore do not affect the conclusions of this work.
This is supported by the intrinsic lineshapes for $a_{c} = 4.5$ and 6.0 fm which do not present such numerical discontinuities and are almost identical, see Fig. \ref{fig:Comparison_FitMetricsAndWidths_ChannelRadiusDependence_cropped}(d).
Furthermore, it is observed that the observed lineshapes in Fig. \ref{fig:Comparison_FitMetricsAndWidths_ChannelRadiusDependence_cropped}(e), which include a Gaussian convolution, are in significantly better agreement than the $1\sigma$ uncertainty of the data.
Nevertheless, the optimal fits presented in the primary analysis in Section \ref{subsec:The primary analysis of inclusive spectra} were checked to not exhibit such numerical discontinuities in the range of the $3_{1}^{-}$ resonance.
For future studies, an alternative algorithm detailed is planned to be implemented for lower-energy resonances where data may warrant such precision \cite{MICHEL2007232}.
% The fit results at different channel radii correspond to the fit conditions of case (4) (see Table \ref{fig:Comparison_AnalysisApproximations_new_cropped}), with the exception of $a_{c} = 4.7$ fm in panel (e) which includes the Landau energy loss of case (5).
For completeness, the fits in Fig. \ref{fig:Comparison_FitMetricsAndWidths_ChannelRadiusDependence_cropped}(d) employing $a_{c} = 4.5$, 5.1 and 6.0 fm all systematically underestimate the data between $E_{x} = 9.61$ to 9.65 MeV.
This is because these fits employ the conditions of case (4) (see Table \ref{tab:FitResults_IsolatedAnalyses_PR146_16deg}) which closely mimics the fitting procedure in Ref. \cite{PhysRevC.87.057307} which ignores target-related energy loss effects.
By including a Landau energy loss in the experimental response, corresponding to case (5), this discrepancy is drastically reduced.

%================================================================================================
\subsubsection{\label{subsubsec:Isolated analysis of the 14C_p_t_12C (theta_lab = 21 deg) data}Isolated analysis of $\mathrm{^{14}C}(p, t)\mathrm{^{12}C}$ ($\theta_{\textrm{lab}} = 21^{\circ}$) data}

\begin{figure}[hbtp]
\includegraphics[width=\columnwidth]{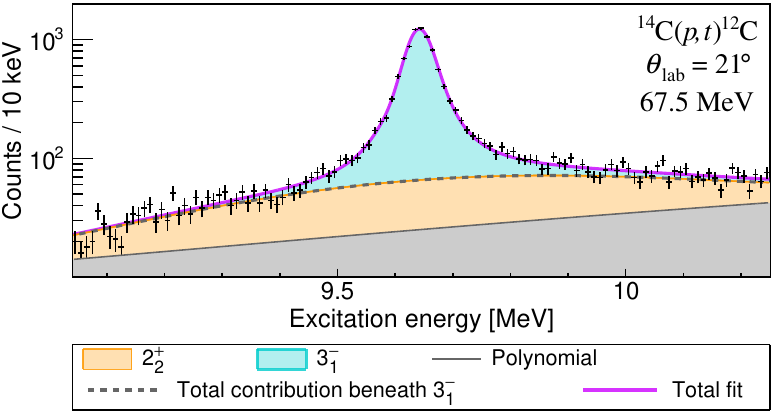}% Here is how to import EPS art
\caption{ \label{fig:AssembleFittedSpectra_IsolatedAnalyses_3minus_letter_only_14C_p_t_12C_cropped}
(Color online)
Decomposition of the fit for case (4) in Table \ref{tab:FitResults_IsolatedAnalyses_PR146_16deg}.
The $3_{1}^{-}$ resonance is superimposed on the total contribution of the $2_{2}^{+}$ resonance (parameterized with a Gaussian lineshape) and a polynomial background.
The reaction, measurement angle and beam energy are indicated.
}
\end{figure}

The same systematic fitting procedure detailed in Section \ref{subsubsec:Isolated analysis of 12C_p_p_12C data} was repeated for the $\mathrm{^{14}C}(p, t)\mathrm{^{12}C}$ ($\theta_{\textrm{lab}} = 21^{\circ}$) data which exhibits the highest resolution from the considered data of 32(1) keV FWHM.
Furthermore, the $\mathrm{^{14}C}(p, t)\mathrm{^{12}C}$ ($\theta_{\textrm{lab}} = 21^{\circ}$) data exhibits the most selective population for the $3_{1}^{-}$ resonance relative to the surrounding $2_{2}^{+}$ and monopole contributions.
The same systematic trends were observed; Fig. \ref{fig:AssembleFittedSpectra_IsolatedAnalyses_3minus_letter_only_14C_p_t_12C_cropped} presents the corresponding fit under the same analysis conditions as Ref. \cite{PhysRevC.87.057307}: case (4).
Case (4) for $\mathrm{^{14}C}(p, t)\mathrm{^{12}C}$ ($\theta_{\textrm{lab}} = 21^{\circ}$) similarly yields $\Gamma(E_{r}) = 48(2)$ keV with $\Gamma_{\textrm{obs}}(E_{r}) = 38(2)$ keV.
The results of this isolated analysis for the $\mathrm{^{14}C}(p, t)\mathrm{^{12}C}$ ($\theta_{\textrm{lab}} = 21^{\circ}$) data mirror those from the isolated analysis of the $\mathrm{^{12}C}(p, p^{\prime})\mathrm{^{12}C}$ data.
This indicates that the approximations employed in Ref. \cite{PhysRevC.105.024308} are all appropriate and that the corresponding observed width of $\Gamma_{\textrm{obs}}(E_{r}) = 39(4)$ keV (converted in this work) should be included in future evaluations.

%------------------------------------------------
%------------------------------------------------

%================================================================================================
\subsection{\label{subsec:Assessing the $3_{1}^{-}$ total width from Alcorta et al.}Assessing the $3_{1}^{-}$ total width from Ref. \cite{PhysRevC.86.064306}} 

\begin{figure}[hbtp]
\includegraphics[width=\columnwidth]{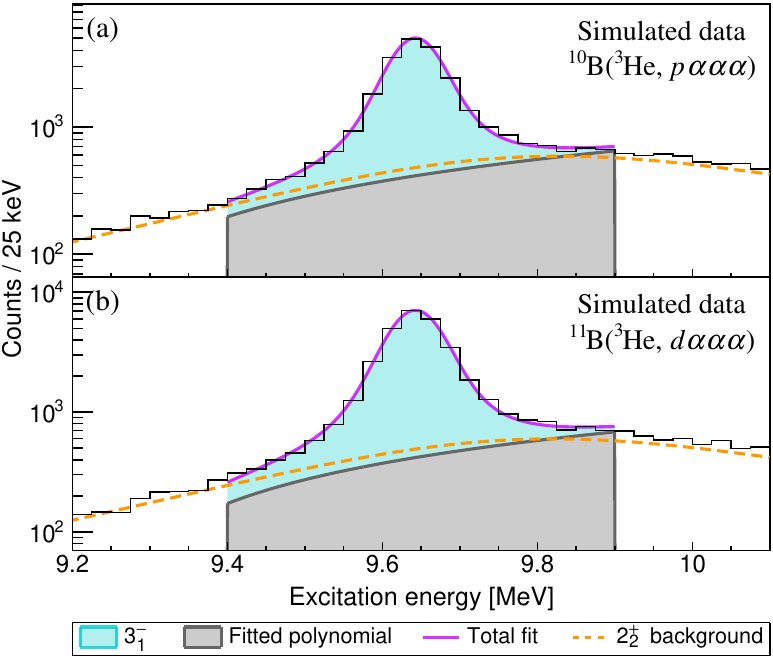}% Here is how to import EPS art
\caption{ \label{fig:AssembleFittedSpectra_CorrectForAlcortaStudy_letter_cropped}
(Color online) Panels (a) and (b) present simulated spectra which mimic the relevant data in Figs. 1(a) and 2(a) in Ref. \cite{PhysRevC.86.064306}, respectively.
The purpose of this analysis is to estimate the unaccounted-for systematic error in Ref. \cite{PhysRevC.86.064306}. See text for details.
}
\end{figure}

The fit of the $3_{1}^{-}$ resonance in Ref. \cite{PhysRevC.86.064306} does not employ any \textbf{R}-matrix-derived lineshapes.
Instead, the fit analysis in Ref. \cite{PhysRevC.86.064306} was performed with a standard/symmetric Lorentzian (that does not capture the intrinsic asymmetry of the $3_{1}^{-}$ resonance) which was convoluted with a Gaussian experimental resolution.
% for the $3_{1}^{-}$ resonance.
% This symmetric Lorentzian, which does not capture the intrinsic asymmetry of the $3_{1}^{-}$ resonance, was convoluted with a Gaussian experimental resolution.
The analysis in Section \ref{subsec:Assessing the $3_{1}^{-}$ total width from Kokalova et al.} and Table \ref{tab:FitResults_IsolatedAnalyses_PR146_16deg} shows that such a Lorentzian lineshape may provide a reasonable estimation of the total width, albeit possibly with a poorer quality fit.
However, Ref. \cite{PhysRevC.86.064306} does not implement the $2_{2}^{+}$ background, which has been observed to only improve the fit, but affect the extracted $3_{1}^{-}$ total width.
The simplified background and symmetric $3_{1}^{-}$ lineshape implemented by Ref. \cite{PhysRevC.86.064306} thus introduces an unaccounted-for systematic error which is dependent on the relative population of the $2_{2}^{+}$ and $3_{1}^{-}$ resonances.
In this work, this systematic error is conservatively estimated by simulating the $\mathrm{^{10}B}(\mathrm{^{3}He}, p\alpha\alpha\alpha)$ and $\mathrm{^{11}B}(\mathrm{^{3}He}, d\alpha\alpha\alpha)$ spectra which correspond to Figs. 1(a) and 2(a) in Ref. \cite{PhysRevC.86.064306}, respectively (see Fig. \ref{fig:AssembleFittedSpectra_CorrectForAlcortaStudy_letter_cropped}).
%------------------------------------------------
%------------------------------------------------
The simulated data was generated using two isolated (single-channel) levels: the $3_{1}^{-}$ resonance with $\Gamma_{\textrm{obs}}(E_{r}) = 38$ keV at $E_{r} = 9.641$ MeV and the $2_{2}^{+}$ resonance with $\Gamma_{\textrm{obs}}(E_{r}) = 850$ keV at $E_{r} = 9.870$ MeV (corresponding to the current ENSDF evaluation \cite{KELLEY201771}).
To roughly mimic the relevant range of the $\mathrm{^{10}B}(\mathrm{^{3}He}, p\alpha\alpha\alpha)$ spectrum, the $2_{2}^{+}$ and $3_{1}^{-}$ resonances were populated with $2.8 \times 10^{4}$ and $2.0 \times 10^{4}$ counts, respectively.
Similarly for the $\mathrm{^{10}B}(\mathrm{^{3}He}, p\alpha\alpha\alpha)$ spectrum, the $2_{2}^{+}$ and $3_{1}^{-}$ resonances are populated with $2.8 \times 10^{4}$ and $3.0 \times 10^{4}$ counts, respectively.
The simulated $\mathrm{^{10}B}(\mathrm{^{3}He}, p\alpha\alpha\alpha)$ and $\mathrm{^{10}B}(\mathrm{^{3}He}, p\alpha\alpha\alpha)$ spectra were convoluted with 55 and 60 keV Gaussian resolutions, respectively (corresponding to the best resolutions for each measurement reported in Ref. \cite{PhysRevC.86.064306}).
The fits in Figs. \ref{fig:AssembleFittedSpectra_CorrectForAlcortaStudy_letter_cropped} were performed with symmetric Lorentzian functions for the $3_{1}^{-}$ resonance (convoluted with the aforementioned experimental resolutions) and second-order polynomials.
The total observed width of the $3_{1}^{-}$ resonance were extracted as 45 and 43 keV for the simulated $\mathrm{^{10}B}(\mathrm{^{3}He}, p\alpha\alpha\alpha)$ and $\mathrm{^{11}B}(\mathrm{^{3}He}, d\alpha\alpha\alpha)$ spectra, respectively.
These widths are systematically higher than the simulated observed total width of 38 keV; a similar difference is observed between the optimised observed width of $\Gamma_{\textrm{obs}}(E_{r}) = 38(2)$ reported this work and the original width of $\Gamma = 43(4)$ reported in Ref. \cite{PhysRevC.86.064306}.
The systematic error within the $3_{1}^{-}$ total width extraction of Ref. \cite{PhysRevC.86.064306} is therefore conservatively estimated to be $45 - 38 = 7$ keV.
% , which can be added in quadrature to the original width of $\Gamma = 43(4)$ reported by Ref. \cite{PhysRevC.86.064306}.
% Adding this systematic error in quadrature yields a modified value of $\Gamma = 43(8)$ keV, which is recommended for future evaluations of the $3_{1}^{-}$ total width that consider Ref. \cite{PhysRevC.86.064306}.
Adding this systematic error in quadrature yields a modified value of $\Gamma = 43(8)$ keV, which is recommended for future evaluations which consider Ref. \cite{PhysRevC.86.064306}.
% It is recommended that the ed value of $\Gamma = 43(8)$ keV (based on the result from Ref. \cite{PhysRevC.86.064306}) be included in future evaluations.

% Width of $3_{1}^{-}$ resonance
% $\mathrm{^{10}B}(\mathrm{^{3}He}, p\alpha\alpha\alpha)$: 45 keV
% $\mathrm{^{11}B}(\mathrm{^{3}He}, d\alpha\alpha\alpha)$: 43 keV

%================================================================================================
\subsection{\label{subsec:Assessing the $3_{1}^{-}$ total width from Browne et al.}Assessing the $3_{1}^{-}$ total width from Ref. \cite{PhysRev.125.992}}

\begin{figure}[hbtp]
\includegraphics[width=\columnwidth]{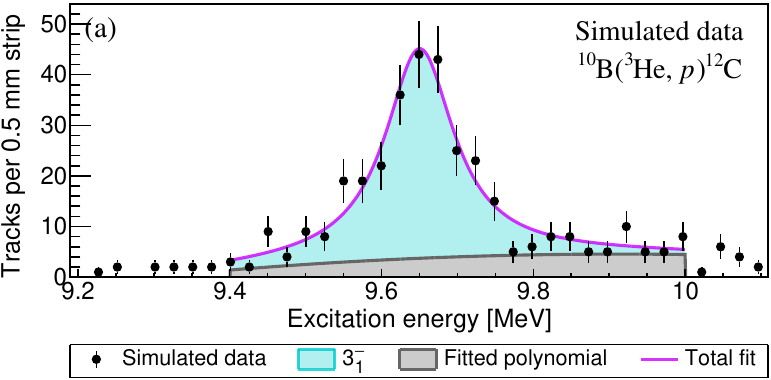}% Here is how to import EPS art
\caption{ \label{fig:AssembleFittedSpectra_CorrectForBrowneStudy_letter_cropped}
(Color online) A simulated spectrum which mimics the nuclear-emulsion-plate data in Fig. 2 of Ref. \cite{PhysRev.125.992}.
The purpose of this analysis is to estimate the unaccounted-for systematic error in Ref. \cite{PhysRev.125.992}.
See text for details.
}
\end{figure}

In Ref. \cite{PhysRev.125.992}, the width of the $3_{1}^{-}$ resonance was obtained by fitting the spectra corresponding to the $\mathrm{^{10}B}(\mathrm{^{3}He}, p)\mathrm{^{12}C}$ reaction.
For the width of the $3_{1}^{-}$ resonance, the corresponding peak was directly fitted with a Lorentzian lineshape (with no Gaussian/experimental convolution).
The resultant FWHM of the Lorentzian peak was assumed to correspond to the intrinsic FWHM of the $3_{1}^{-}$ resonance added in quadrature with the experimental resolution. 
This basic approximation yields a systematic error: this is tested by convoluting a $\Gamma_{\textrm{obs}}(E_{r}) = 38$ keV $3_{1}^{-}$ resonance with a Gaussian resolution of $\approx 100$ keV FWHM observed in Fig. 2 of Ref. \cite{PhysRev.125.992}.
The numerical convolution yields an FWHM of 122 keV whilst adding the widths in quadrature yields an FHWM 107 keV.
Another source of systematic error is the fitting methodology in Ref. \cite{PhysRev.125.992} which employed a symmetric Lorentzian with no Gaussian convolution component for the experimental resolution.
Finally, the underlying $2_{2}^{+}$ resonance is not accounted for in Ref. \cite{PhysRev.125.992}; a consequence of Ref. \cite{PhysRev.125.992} being performed roughly half a century before the existence of the underlying $2_{2}^{+}$ resonance was confirmed \cite{MItoh_2004, PhysRevC.80.041303, PhysRevC.84.054308, PhysRevC.86.034320, PhysRevLett.110.152502}.
To estimate the overall unaccounted-for systematic error from these intertwined approximations, the $\mathrm{^{10}B}(\mathrm{^{3}He}, p)\mathrm{^{12}C}$ spectrum in Fig. 2 of Ref. \cite{PhysRev.125.992} has been simulated in Fig. \ref{fig:AssembleFittedSpectra_CorrectForBrowneStudy_letter_cropped} of this work.
As in Section \ref{subsec:Assessing the $3_{1}^{-}$ total width from Alcorta et al.}, the simulated data employed two isolated (single-channel) levels: the $3_{1}^{-}$ resonance with $\Gamma_{\textrm{obs}}(E_{r}) = 38$ keV at $E_{r} = 9.641$ MeV and the $2_{2}^{+}$ resonance with $\Gamma_{\textrm{obs}}(E_{r}) = 850$ keV at $E_{r} = 9.870$ MeV (corresponding to the current ENSDF evaluation \cite{KELLEY201771}).
To roughly mimic the $\mathrm{^{10}B}(\mathrm{^{3}He}, p)\mathrm{^{12}C}$ data in Ref. \cite{PhysRev.125.992}, the $2_{2}^{+}$ and $3_{1}^{-}$ resonances were each populated with 250 counts.
The fitted Lorentzian yielded a total width of 110 keV and following the procedure in Ref. \cite{PhysRev.125.992} of removing the instrumental width by ``taking the square root of the difference of the squares'' yields $\sqrt{110^{2} - 100^{2}} \approx 47$ keV, producing a total systematic error of approximately $47 - 38 = 9$ keV.
This systematic error is added in quadrature to the width reported in Ref. \cite{PhysRev.125.992} to yield a modified value of $\Gamma = 36(11)$ keV, which is recommended for future evaluations which consider Ref. \cite{PhysRev.125.992}.

%------------------------------------------------
%------------------------------------------------

%================================================================================================
\section{\label{sec:Discussion}DISCUSSION}

The primary analysis in this work (Section \ref{subsec:The primary analysis of inclusive spectra}) yields $\Gamma(E_{r}) = 46(2)$ keV with $\Gamma_{\textrm{obs}}(E_{r}) = 38(2)$ keV for the $3_{1}^{-}$ resonance in $\mathrm{^{12}C}$.
The observed width from this work is significantly smaller than the current ENSDF average of 46(3) keV which has been employed in Ref. \cite{TSUMURA2021136283}.
A meta-analysis was performed on the results currently considered in the current ENSDF evaluation \cite{PhysRev.125.992, PhysRevC.86.064306, PhysRevC.87.057307} (see sections \ref{subsec:Assessing the $3_{1}^{-}$ total width from Kokalova et al.}, \ref{subsec:Assessing the $3_{1}^{-}$ total width from Alcorta et al.} and \ref{subsec:Assessing the $3_{1}^{-}$ total width from Browne et al.}).
In this work, the misstated formal total width in Ref. \cite{PhysRevC.87.057307} was converted to the appropriate observed total width for the ENSDF.
The unaccounted-for uncertainties in Refs. \cite{PhysRev.125.992, PhysRevC.86.064306, PhysRevC.87.057307} were estimated to yield modified values which we recommend for future evaluations.
A quantitative assessment of Ref. \cite{PhysRev.104.1059} was deemed unfeasible within the scope of this work as the experimental methodology in Ref. \cite{PhysRev.104.1059} is significantly different from this work.
However, the systematics in sections \ref{subsec:Assessing the $3_{1}^{-}$ total width from Kokalova et al.}, \ref{subsec:Assessing the $3_{1}^{-}$ total width from Alcorta et al.} and \ref{subsec:Assessing the $3_{1}^{-}$ total width from Browne et al.} show that not accounting for the asymmetric background from the underlying $2_{1}^{+}$ resonance gives rise to a systematic error.
Therefore, until the analysis of Ref. \cite{PhysRev.104.1059} is appropriately reviewed, we recommend the associated result be omitted from future evaluations.
A summary of the recommended results for the $3_{1}^{-}$ resonance in $\mathrm{^{12}C}$ is given in Table \ref{tab:RecommendedDataForNNDCEvaluationOfWidth}.
To reiterate: $\Gamma_{\textrm{obs}}(E_{r})$ and $\Gamma_{\textrm{FWHM}}$ are equivalent for the $3_{1}^{-}$ resonance, see Section \ref{subsec:Assessing the $3_{1}^{-}$ total width from Kokalova et al.}.

%================================================================================================
\begin{table}[btp]%The best place to locate the table environment is directly after its first reference in text
\caption{\label{tab:RecommendedDataForNNDCEvaluationOfWidth}%
The results for the $3_{1}^{-}$ resonance in $\mathrm{^{12}C}$ which are recommended for future ENSDF evaluations.
For analyses which employed the \textbf{R}-matrix formalism, $a_{c}$ denotes the channel radius.
% The formal ($\Gamma(E_{r})$), observed ($\Gamma_{\textrm{obs}}$) and FWHM ($\Gamma_{\textrm{FWHM}}$) widths are summarised, with the total uncertainties modified in this work.
The formal, observed and FWHM widths are summarised, respectively denoted as $\Gamma(E_{r})$, $\Gamma_{\textrm{obs}}$ and $\Gamma_{\textrm{FWHM}}$.
For the physical total widths of previous works (i.e. $\Gamma_{\textrm{obs}}$ and $\Gamma_{\textrm{FWHM}}$), the total uncertainties were modified in this work.
% $\Gamma(E_{r})$ is the formal total width (see Eq. \ref{eq:FormalTotalWidth}).
% $\Gamma_{\textrm{obs}}$ is the observed total width (see Eq. \ref{eq:ObservedTotalWidth}). 
% $\Gamma_{\textrm{FWHM}}$ is the full width at half maximum of the intrinsic lineshape.
}
\begin{ruledtabular}
%================================================
% \begin{tabular}{l c D{,}{}{3.3} D{,}{}{3.3}}
% \multicolumn{1}{c}{Ref.}    & $\Gamma(E_{r})$ & \multicolumn{1}{c}{$\Gamma_{\textrm{obs}}(E_{r})$} & \multicolumn{1}{c}{$\Gamma_{\textrm{FWHM}}$} \\
%          & {[keV]} & \multicolumn{1}{c}{[keV]} & \multicolumn{1}{c}{[keV]} \\ [0.3ex] \hline \\ [-2.3ex]
% \citeauthor{PhysRev.125.992} \cite{PhysRev.125.992}      &---& \multicolumn{1}{c}{---} & 36,(11)\\ [0.3ex] \hline \\ [-2.3ex]
% \citeauthor{PhysRevC.86.064306} \cite{PhysRevC.86.064306}   &---& \multicolumn{1}{c}{---} & 43,(8) \\ [0.3ex] \hline \\ [-2.3ex]
% \citeauthor{PhysRevC.87.057307} \cite{PhysRevC.87.057307}   & 48(2) & 39,(4)\footnotemark[1] &39,(4)\footnotemark[1] \\ [0.3ex] \hline \\ [-2.3ex]
% This work   & 46(2) & 38,(2) &38,(2) \\
% \end{tabular}
%================================================
\begin{tabular}{l D{,}{}{3.3} c D{,}{}{3.3} D{,}{}{3.3}}
\multicolumn{1}{c}{Ref.}    & \multicolumn{1}{c}{$a_{c}$} & $\Gamma(E_{r})$ & \multicolumn{1}{c}{$\Gamma_{\textrm{obs}}(E_{r})$} & \multicolumn{1}{c}{$\Gamma_{\textrm{FWHM}}$} \\
         & \multicolumn{1}{c}{[fm]} & {[keV]} & \multicolumn{1}{c}{[keV]} & \multicolumn{1}{c}{[keV]} \\ [0.3ex] \hline \\ [-2.3ex]
\citeauthor{PhysRev.125.992} \cite{PhysRev.125.992}      &\multicolumn{1}{c}{---}&---&  \multicolumn{1}{c}{---} & 36,(11)\\ [0.3ex] \hline \\ [-2.3ex]
\citeauthor{PhysRevC.86.064306} \cite{PhysRevC.86.064306}   &\multicolumn{1}{c}{---}&---& \multicolumn{1}{c}{---} & 43,(8) \\ [0.3ex] \hline \\ [-2.3ex]
\citeauthor{PhysRevC.87.057307} \cite{PhysRevC.87.057307}   & 4,.7\footnotemark[1] & 48(2) & 39,(4)\footnotemark[2] &39,(4)\footnotemark[2] \\ [0.3ex] \hline \\ [-2.3ex]
This work   & 4,.8 & 46(2) & 38,(2) &38,(2) \\
\end{tabular}
%================================================
\end{ruledtabular}
\footnotetext[1]{The exact channel radius being $a_{c} = 1.3(4^{1/3} + 8^{1/3})$ fm \cite{PhysRevC.87.057307}.}
\footnotetext[2]{Not reported in Ref. \cite{PhysRevC.87.057307}; converted from $\Gamma(E_{r})$ in this work.
% The reported uncertainties include systematic errors which account for the choice in background (see text for details).
}
\end{table}
%================================================================================================

The nuclear structure of the $3_{1}^{-}$ state also serves as an important test of theoretical models for $\mathrm{^{12}C}$.
The $3_{1}^{-}$ state has been suggested as a candidate for the $K^{\pi} = 3^{-}$ bandhead \cite{PhysRevLett.81.5291, RevModPhys.90.035004} and is understood to exhibit significant $\alpha$-cluster structure \cite{PhysRevC.87.057307} with a dominant $\alpha_{0}$ decay mode \cite{PhysRevC.76.034320, PhysRevC.85.037603, PhysRevC.86.064306}.
Interestingly, a recent ab initio calculation using nuclear lattice effective field theory has predicted the $3_{1}^{-}$ and $2_{1}^{+}$ states to exhibit equilateral triangle symmetry for the constituent alpha clusters \cite{S_Shihang_2023}.
The degree of clustering of a resonance for a particular decay partition can be gleaned from the associated Wigner ratio given by $\theta^{2} = \gamma^{2}/\gamma_{W}^{2}$, where $\gamma_{W}^{2} = 3 \hbar^{2}/2\mu a^{2}$, with $\mu$ and $a_{c}$ being the reduced mass and channel radius, respectively \cite{PhysRev.87.123}.
The optimal fit in Fig. \ref{fig:AssembleFittedSpectra_GlobalAnalyses_3minus_letter_cropped} yields an $\alpha_{0}$ Wigner ratio of $\approx30\%$, which indicates a large degree of clustering/preformation for the $\mathrm{^{12}C} \rightarrow \mathrm{^{8}Be}(0_{\textrm{g.s.}}^{+}) + \alpha$ $(l=3)$ partition.
This Wigner ratio, which is highly dependent on the channel radius, is in agreement with that reported in Ref. \cite{PhysRevC.87.057307} as Kokalova \textit{et al.} employed an $\alpha_{0}$ channel radius of $\approx 4.7$ fm, which is very similar to the optimized 4.8 fm channel radius in this work.
The total observed width of $\Gamma_{\textrm{obs}}(E_{r}) = 38(2)$ is in reasonable agreement with the theoretical prediction of 30 keV by Uegaki \textit{et al.} \cite{GCM_UegakiE}, with the total width of 68 keV predicted by {\'A}lvarez-Rodr{\'i}guez \textit{et al.} being somewhat larger \cite{Alvarez-Rodriguez2007}.

\section{\label{sec:Conclusions}CONCLUSIONS}

In this work, the physical total width of the $3_{1}^{-}$ resonance in $\mathrm{^{12}C}$ was studied with a global analysis of high-resolution spectra populated with direct reactions.
A simultaneous fit analysis yielded a formal total width of $\Gamma(E_{r}) = 46(2)$ keV and an observed total width of $\Gamma_{\textrm{obs}}(E_{r}) = 38(2)$ keV.
This result is significantly discrepant with the current ENSDF average of 46(3) keV for the total width of the $3_{1}^{-}$ resonance \cite{KELLEY201771}.
To investigate this inconsistency, a meta-analysis was performed on all previous results currently considered in the current ENSDF evaluation for the $3_{1}^{-}$ resonance \cite{KELLEY201771} (with the exception of Ref. \cite{PhysRev.104.1059}).
It was concluded that all these previous results \cite{KELLEY201771} contain unaccounted-for systematic errors, with a single study reporting a misstated total width \cite{PhysRevC.87.057307}.
% In particular, Ref. \cite{PhysRevC.87.057307} misstated the model-dependent formal total width of $\Gamma(E_{r}) = 48(2)$ (relevant to \textbf{R}-matrix theory) as the physical (or observed) total width of the $3_{1}^{-}$ resonance.
An uncertainty-weighted average of the recommend observed (physical) total widths for the $3_{1}^{-}$ resonance yields a total width of $\Gamma_{\textrm{FWHM}}$ = 38(2) keV (see Table \ref{tab:RecommendedDataForNNDCEvaluationOfWidth}).
This physical width is recommended for future evaluations of the observed total radiative width for the $3_{1}^{-}$ resonance and its contribution to the high-temperature triple-$\alpha$ reaction rate.

\begin{acknowledgments}

This work is based on the research supported in part by the National Research Foundation of South Africa (Grant Numbers: 85509, 86052, 118846, 90741).
The authors acknowledge the accelerator staff of iThemba LABS for providing excellent beams.
The computations were performed on resources provided by UNINETT Sigma2 - the National Infrastructure for High Performance Computing and Data Storage in Norway.
K.~C.~W.~Li would like to thank H.~O.~U.~Fynbo, R.~J.~deBoer, C.~R.~Brune, A.~C.~Larsen and A.~S.~Voyles for useful discussions, as well as S.~Basunia and J.~H.~Kelley for communicating details of the ENSDF evaluations.
% K.~C.~W.~Li would like to thank H.~O.~U.~Fynbo, R.~J.~deBoer and C.~R.~Brune for useful discussions.
% The authors would like to thank S.~Basunia and J.~H.~Kelley 
\end{acknowledgments}

% \section{A little more on appendixes}

% Observe that this appendix was started by using
% \begin{verbatim}
% \section{A little more on appendixes}
% \end{verbatim}

% Note the equation number in an appendix:
% \begin{equation}
% E=mc^2.
% \end{equation}

% \nocite{*}

\bibliography{12C_2021}% Produces the bibliography via BibTeX.

\end{document}